\documentclass[twocolumn]{style/aastex63}
\usepackage{amsmath}
\usepackage{graphicx}
\usepackage{natbib}
\usepackage[scaled]{helvet}
\usepackage{epsfig}
\usepackage{url}
\bibpunct{(}{)}{;}{a}{}{,}
\newcommand{\kms}{\,km\,s$^{-1}$}
\usepackage{textcomp}
\usepackage{mathpazo}
\usepackage{xspace}
\usepackage{hyperref}
\usepackage{style/apjfonts}
\usepackage{mathrsfs}
\usepackage{verbatim}

\usepackage{color}
\usepackage[normalem]{ulem}

\newcommand{\bjdtdb}{\ensuremath{\rm {BJD_{TDB}}}}
\newcommand{\feh}{\ensuremath{\left[{\rm Fe}/{\rm H}\right]}}

\newcommand{\teff}{\ensuremath{T_{\rm eff}}\xspace}
\newcommand{\logg}{\ensuremath{\log g}}

\newcommand{\msun}{\ensuremath{\,M_\Sun}}
\newcommand{\rsun}{\ensuremath{\,R_\Sun}}
\newcommand{\lsun}{\ensuremath{\,L_\Sun}}
\newcommand{\mj}{\ensuremath{\,M_{\rm J}}}
\newcommand{\rj}{\ensuremath{\,R_{\rm J}}}

\newcommand{\ms}{\,m\,s$^{-1}$}
\newcommand{\thisstarone}{TOI-558\xspace}
\newcommand{\thisstartwo}{TOI-559\xspace}

\newcommand{\mstar}{\ensuremath{M_{*}}}
\newcommand{\rstar}{\ensuremath{R_{*}}}

\newcommand{\be}{\begin{equation}}
\newcommand{\ee}{\end{equation}}

\newcommand{\tess}{{\it TESS}}
\newcommand{\TESS}{{\it TESS}}
\usepackage{lineno}

\begin{document}

\title{Two Massive Jupiters in Eccentric Orbits from the TESS Full Frame Images}

\newcommand{\cfa}{Center for Astrophysics \textbar \ Harvard \& Smithsonian, 60 Garden St, Cambridge, MA 02138, USA}
\newcommand{\msu}{Department of Physics and Astronomy, Michigan State University, East Lansing, MI 48824, USA}
\newcommand{\umich}{Astronomy Department, University of Michigan, 1085 S University Avenue, Ann Arbor, MI 48109, USA}
\newcommand{\utaustin}{Department of Astronomy, The University of Texas at Austin, Austin, TX 78712, USA}
\newcommand{\MIT}{Department of Physics and Kavli Institute for Astrophysics and Space Research, Massachusetts Institute of Technology, Cambridge, MA 02139, USA}
\newcommand{\MITEPS}{Department of Earth, Atmospheric and Planetary Sciences, Massachusetts Institute of Technology,  Cambridge,  MA 02139, USA}
\newcommand{\uflorida}{Department of Astronomy, University of Florida, 211 Bryant Space Science Center, Gainesville, FL, 32611, USA}
\newcommand{\riverside}{Department of Earth Sciences, University of California,
Riverside, CA 92521, USA}
\newcommand{\usq}{Centre for Astrophysics, University of Southern Queensland, West Street, Toowoomba, QLD 4350, Australia}
\newcommand{\ames}{NASA Ames Research Center, Moffett Field, CA, 94035, USA}
\newcommand{\geneva}{Observatoire de l’Universit\'e de Gen\`eve, 51 chemin des Maillettes,
1290 Versoix, Switzerland}
\newcommand{\uw}{Astronomy Department, University of Washington, Seattle, WA 98195 USA}
\newcommand{\warwick}{Deptartment of Physics, University of Warwick, Gibbet Hill Road, Coventry CV4 7AL, UK}
\newcommand{\warwickceh}{Centre for Exoplanets and Habitability, University of Warwick, Gibbet Hill Road, Coventry CV4 7AL, UK}
\newcommand{\princeton}{Department of Astrophysical Sciences, Princeton University, 4 Ivy Lane, Princeton, NJ, 08544, USA}
\newcommand{\liege}{Space Sciences, Technologies and Astrophysics Research (STAR) Institute, Universit\'e de Li\`ege, 19C All\'ee du 6 Ao\^ut, 4000 Li\`ege, Belgium}
\newcommand{\vanderbilt}{Department of Physics and Astronomy, Vanderbilt University, Nashville, TN 37235, USA}
\newcommand{\fisk}{Department of Physics, Fisk University, 1000 17th Avenue North, Nashville, TN 37208, USA}
\newcommand{\columbia}{Department of Astronomy, Columbia University, 550 West 120th Street, New York, NY 10027, USA}
\newcommand{\toronto}{Dunlap Institute for Astronomy and Astrophysics, University of Toronto, Ontario M5S 3H4, Canada}
\newcommand{\unc}{Department of Physics and Astronomy, University of North Carolina at Chapel Hill, Chapel Hill, NC 27599, USA}
\newcommand{\iac}{Instituto de Astrof\'isica de Canarias (IAC), E-38205 La Laguna, Tenerife, Spain}
\newcommand{\lalaguna}{Departamento de Astrof\'isica, Universidad de La Laguna (ULL), E-38206 La Laguna, Tenerife, Spain}
\newcommand{\louisville}{Department of Physics and Astronomy, University of Louisville, Louisville, KY 40292, USA}
\newcommand{\aavso}{American Association of Variable Star Observers, 49 Bay State Road, Cambridge, MA 02138, USA}
\newcommand{\utokyo}{The University of Tokyo, 7-3-1 Hongo, Bunky\={o}, Tokyo 113-8654, Japan}
\newcommand{\naoj}{National Astronomical Observatory of Japan, 2-21-1 Osawa, Mitaka, Tokyo 181-8588, Japan}
\newcommand{\jstpresto}{JST, PRESTO, 7-3-1 Hongo, Bunkyo-ku, Tokyo 113-0033, Japan}
\newcommand{\astrobiojapan}{Astrobiology Center, 2-21-1 Osawa, Mitaka, Tokyo 181-8588, Japan}
\newcommand{\ctio}{Cerro Tololo Inter-American Observatory, Casilla 603, La Serena, Chile}
\newcommand{\nexsci}{Caltech IPAC -- NASA Exoplanet Science Institute 1200 E. California Ave, Pasadena, CA 91125, USA}
\newcommand{\ucsc}{Department of Astronomy and Astrophysics, University of
California, Santa Cruz, CA 95064, USA}
\newcommand{\gsfc}{Exoplanets and Stellar Astrophysics Laboratory, Code 667, NASA Goddard Space Flight Center, Greenbelt, MD 20771, USA}
\newcommand{\sgtinc}{SGT, Inc./NASA AMES Research Center, Mailstop 269-3, Bldg T35C, P.O. Box 1, Moffett Field, CA 94035, USA}
\newcommand{\chile}{Center of Astro-Engineering UC, Pontificia Universidad Cat\'olica de Chile, Av. Vicu\~{n}a Mackenna 4860, 7820436 Macul, Santiago, Chile}
\newcommand{\Pontificia}{Instituto de Astrof\'isica, Pontificia Universidad Cat\'olica de Chile, Av.\ Vicu\~na Mackenna 4860, Macul, Santiago, Chile}
\newcommand{\Millennium}{Millennium Institute for Astrophysics, Chile}
\newcommand{\maxplank}{Max-Planck-Institut f\"ur Astronomie, K\"onigstuhl 17, Heidelberg 69117, Germany}
\newcommand{\utdallas}{Department of Physics, The University of Texas at Dallas, 800 West
Campbell Road, Richardson, TX 75080-3021 USA}
\newcommand{\MauryLewin}{Maury Lewin Astronomical Observatory, Glendora, CA 91741, USA}
\newcommand{\umbc}{University of Maryland, Baltimore County, 1000 Hilltop Circle, Baltimore, MD 21250, USA}
\newcommand{\osu}{Department of Astronomy, The Ohio State University, 140 West 18th Avenue, Columbus, OH 43210, USA}
\newcommand{\MITAA}{Department of Aeronautics and Astronautics, MIT, 77 Massachusetts Avenue, Cambridge, MA 02139, USA}
\newcommand{\openu}{School of Physical Sciences, The Open University, Milton Keynes MK7 6AA, UK}
\newcommand{\swarthmore}{Department of Physics and Astronomy, Swarthmore College, Swarthmore, PA 19081, USA}
\newcommand{\seti}{SETI Institute, Mountain View, CA 94043, USA}
\newcommand{\lehigh}{Department of Physics, Lehigh University, 16 Memorial Drive East, Bethlehem, PA 18015, USA}
\newcommand{\utah}{Department of Physics and Astronomy, University of Utah, 115 South 1400 East, Salt Lake City, UT 84112, USA}
\newcommand{\USNA}{Department of Physics, United States Naval Academy, 572C Holloway Rd., Annapolis, MD 21402, USA}
\newcommand{\eplcarnegie}{Earth \& Planets Laboratory, Carnegie Institution for Science, 5241 Broad Branch Road, NW, Washington, DC 20015, USA}
\newcommand{\UPenn}{The University of Pennsylvania, Department of Physics and Astronomy, Philadelphia, PA, 19104, USA}
\newcommand{\montana}{Department of Physics and Astronomy, University of Montana, 32 Campus Drive, No. 1080, Missoula, MT 59812 USA}
\newcommand{\psu}{Department of Astronomy \& Astrophysics, The Pennsylvania State University, 525 Davey Lab, University Park, PA 16802, USA}
\newcommand{\psust}{Center for Exoplanets and Habitable Worlds, The Pennsylvania State University, 525 Davey Lab, University Park, PA 16802, USA}
\newcommand{\Kutztown}{Department of Physical Sciences, Kutztown University, Kutztown, PA 19530, USA}
\newcommand{\udel}{Department of Physics \& Astronomy, University of Delaware, Newark, DE 19716, USA}
\newcommand{\Westminster}{Department of Physics, Westminster College, New Wilmington, PA 16172}
\newcommand{\steward}{Department of Astronomy and Steward Observatory, University of Arizona, Tucson, AZ 85721, USA}
\newcommand{\saao}{South African Astronomical Observatory, PO Box 9, Observatory, 7935, Cape Town, South Africa}
\newcommand{\salt}{Southern African Large Telescope, PO Box 9, Observatory, 7935, Cape Town, South Africa}
\newcommand{\ssl}{Societ\`{a} Astronomica Lunae, Italy}
\newcommand{\spot}{Spot Observatory, Nashville, TN 37206, USA}
\newcommand{\txamGP}{George P.\ and Cynthia Woods Mitchell Institute for Fundamental Physics and Astronomy, Texas A\&M University, College Station, TX77843 USA}
\newcommand{\txam}{Department of Physics and Astronomy, Texas A\&M university, College Station, TX 77843 USA}
\newcommand{\wellesley}{Department of Astronomy, Wellesley College, Wellesley, MA 02481, USA}
\newcommand{\byu}{Department of Physics and Astronomy, Brigham Young University, Provo, UT 84602, USA}
\newcommand{\Hazelwood}{Hazelwood Observatory, Churchill, Victoria, Australia}
\newcommand{\pest}{Perth Exoplanet Survey Telescope, Perth, Australia}
\newcommand{\Winer}{Winer Observatory, PO Box 797, Sonoita, AZ 85637, USA}
\newcommand{\icpo}{Ivan Curtis Private Observatory}
\newcommand{\elsauce}{El Sauce Observatory, Chile}
\newcommand{\crow}{Atalaia Group \& CROW Observatory, Portalegre, Portugal}
\newcommand{\dfus}{Dipartimento di Fisica ``E.R.Caianiello'', Universit\`a di Salerno, Via Giovanni Paolo II 132, Fisciano 84084, Italy}
\newcommand{\indfn}{Istituto Nazionale di Fisica Nucleare, Napoli, Italy}
\newcommand{\sotes}{Gabriel Murawski Private Observatory (SOTES)}
\newcommand{\lco}{Las Cumbres Observatory Global Telescope, 6740 Cortona Dr., Suite 102, Goleta, CA 93111, USA}
\newcommand{\ucsb}{Department of Physics, University of California, Santa Barbara, CA 93106-9530, USA}
\newcommand{\carnegie}{The Observatories of the Carnegie Institution for Science, 813 Santa Barbara St., Pasadena, CA 91101, USA}
\newcommand{\wisconsin}{Department of Astronomy, University of Wisconsin-Madison, Madison, WI 53706, USA}
\newcommand{\Patashnick}{Patashnick Voorheesville Observatory, Voorheesville, NY 12186, USA}
\newcommand{\Algonquin}{Algonquin Regional High School, MA, USA}
\newcommand{\crls}{Cambridge Rindge and Latin High School, MA, USA}
\newcommand{\jpl}{Jet Propulsion Laboratory, California Institute of Technology, 4800 Oak Grove Drive, Pasadena, CA 91109, USA}
\newcommand{\stsci}{Space Telescope Science Institute, Baltimore, MD 21218, USA}
\newcommand{\gsfcsellers}{GSFC Sellers Exoplanet Environments Collaboration, NASA Goddard Space Flight Center, Greenbelt, MD 20771 }
\newcommand{\gmu}{George Mason University, 4400 University Drive MS 3F3, Fairfax, VA 22030, USA}
\newcommand{\hawaii}{Institute for Astronomy, University of Hawaii, Maui, HI 96768, USA}
\newcommand{\harvard}{Harvard University, Cambridge, MA 02138, USA}
\newcommand{\ucscchile}{Departamento de Matem\'atica y F\'isica Aplicadas, Universidad Cat\'olica de la Sant\'isima Concepci\'on, Alonso de Rivera 2850, Concepci\'on, Chile}
\newcommand{\brierfield}{Brierfield Observatory, New South Wales, Australia}\newcommand{\curtin}{Curtin Institute of Radio Astronomy, Curtin University, Bentley, Western Australia 6102}
\newcommand{\torres}{\altaffiliation{Juan Carlos Torres Fellow}}
\newcommand{\sagan}{\altaffiliation{NASA Sagan Fellow}}
\newcommand{\bernoulli}{\altaffiliation{Bernoulli fellow}}
\newcommand{\gruber}{\altaffiliation{Gruber fellow}}
\newcommand{\kavli}{\altaffiliation{Kavli Fellow}}
\newcommand{\peg}{\altaffiliation{51 Pegasi b Fellow}}
\newcommand{\pappalardo}{\altaffiliation{Pappalardo Fellow}}
\newcommand{\hubble}{\altaffiliation{NASA Hubble Fellow}}
\newcommand{\nsfgf}{\altaffiliation{National Science Foundation Graduate Research Fellow}}


\correspondingauthor{Mma Ikwut-Ukwa} 
\email{mma.ikwut-ukwa@cfa.harvard.edu}

\author[0000-0002-4404-5505]{Mma Ikwut-Ukwa} 
\affiliation{\cfa}

\author[0000-0001-8812-0565]{Joseph E. Rodriguez} 
\affiliation{\msu}

\author[0000-0002-8964-8377]{Samuel N. Quinn} 
\affiliation{\cfa}

\author[0000-0002-4891-3517]{George Zhou} 
\affiliation{\cfa}

\author[0000-0001-7246-5438]{Andrew Vanderburg} 
\affiliation{\wisconsin}

\author{Asma Ali}
\affiliation{\cfa}
\affiliation{\Algonquin}

\author{Katya Bunten}
\affiliation{\cfa}
\affiliation{\crls}

\author[0000-0003-0395-9869]{B. Scott Gaudi} 
\affiliation{\osu}

\author[0000-0001-9911-7388]{David W. Latham} 
\affiliation{\cfa}

\author[0000-0002-2532-2853]{Steve B.\ Howell} 
\affiliation{\ames}

\author[0000-0003-0918-7484]{Chelsea X. Huang} 
\torres
\affiliation{\MIT}

\author[0000-0001-6637-5401]{Allyson Bieryla} 
\affiliation{\cfa}

\author[0000-0001-6588-9574]{Karen A.\ Collins}
\affiliation{\cfa}

\author[0000-0001-6416-1274]{Theron W. Carmichael}
\affiliation{\harvard}
\affiliation{\cfa}

\author[0000-0003-2935-7196]{Markus Rabus}
\affiliation{\lco}
\affiliation{\ucsb}
\affiliation{\ucscchile}

\author[0000-0003-3773-5142]{Jason D. Eastman} 
\affiliation{\cfa}

\author[0000-0003-2781-3207]{Kevin I.\ Collins}
\affiliation{\gmu}


\author[0000-0001-5603-6895]{Thiam-Guan Tan}
\affiliation{\pest}
\affiliation{\curtin}

\author[0000-0001-8227-1020]{Richard P. Schwarz} 
\affiliation{\Patashnick}

\author[0000-0002-9810-0506]{Gordon Myers}
\affiliation{\aavso}

\author[0000-0003-2163-1437]{Chris Stockdale}
\affiliation{\Hazelwood}

\author[0000-0003-0497-2651]{John F.\ Kielkopf} 
\affiliation{\louisville}

\author[0000-0002-3940-2360]{Don J. Radford}
\affiliation{\brierfield}

\author[0000-0002-0582-1751]{Ryan J. Oelkers} 
\affiliation{\vanderbilt}

\author[0000-0002-4715-9460]{Jon M. Jenkins} 
\affiliation{\ames}

\author{George R. Ricker} 
\affiliation{\MIT}

\author[0000-0002-6892-6948]{Sara Seager}
\affiliation{\MIT}
\affiliation{\MITEPS}
\affiliation{\MITAA}

\author[0000-0001-6763-6562]{Roland K. Vanderspek} 
\affiliation{\MIT}

\author[0000-0002-4265-047X]{Joshua N. Winn}
\affiliation{\princeton}




\author[0000-0002-0040-6815]{Jennifer Burt}
\affiliation{\jpl}

\author[0000-0003-1305-3761]{R. Paul Butler} 
\affiliation{\eplcarnegie}

\author[0000-0002-2830-5661]{Michael L. Calkins} 
\affiliation{\cfa}


\author[0000-0002-5226-787X]{Jeffrey D. Crane} 
\affiliation{\carnegie}


\author[0000-0003-2519-6161]{Crystal~L.~Gnilka}
\affiliation{\ames}

\author[0000-0002-9789-5474]{Gilbert A. Esquerdo} 
\affiliation{\cfa}

\author{William Fong}
\affiliation{\MIT}

\author[0000-0003-0514-1147]{Laura Kreidberg} 
\affiliation{\maxplank}


\author[0000-0003-3594-1823]{Jessica Mink}
\affiliation{\cfa}

\author[0000-0003-1286-5231]{David R. Rodriguez}
\affiliation{\stsci}

\author{Joshua E. Schlieder} 
\affiliation{\gsfc}
\affiliation{\gsfcsellers}


\author{Stephen Shectman} 
\affiliation{\carnegie}

\author[0000-0002-1836-3120]{Avi Shporer}
\affiliation{\MIT}

\author{Johanna Teske} 
\affiliation{\eplcarnegie}

\author[0000-0002-8219-9505]{Eric B. Ting}
\affiliation{\ames}


\author{Jesus Noel Villase{\~ n}or}
\affiliation{\MIT}

\author[0000-0003-4755-584X]{Daniel A. Yahalomi} 
\affiliation{\columbia}
\affiliation{\cfa}

\shorttitle{TOI-558 b \& 559 b}
\shortauthors{Ikwut-Ukwa et al.}

\begin{abstract}
We report the discovery of two short-period massive giant planets from NASA's Transiting Exoplanet Survey Satellite (TESS). Both systems, TOI-558 (TIC 207110080) and TOI-559 (TIC 209459275), were identified from the 30-minute cadence Full Frame Images and confirmed using ground-based photometric and spectroscopic follow-up observations from TESS's Follow-up Observing Program Working Group. We find that TOI-558 b, which transits an F-dwarf (\mstar=$1.349^{+0.064}_{-0.065}$\ \msun, \rstar=$1.496^{+0.042}_{-0.040}$ \rsun, \teff=$6466^{+95}_{-93}$ K, age $1.79^{+0.91}_{-0.73}$ Gyr) with an orbital period of 14.574 days, has a mass of $3.61\pm0.15$ \mj, a radius of $1.086^{+0.041}_{-0.038}$ \rj, and an eccentric (e=$0.300^{+0.022}_{-0.020}$) orbit. TOI-559 b transits a G-dwarf (\mstar=$1.026\pm0.057$ \msun, \rstar=$1.233^{+0.028}_{-0.026}$ \rsun, \teff=$5925^{+85}_{-76}$ K, { age $6.8^{+2.5}_{-2.0}$ Gyr}) in an eccentric (e=$0.151\pm0.011$) 6.984-day orbit with a mass of $6.01^{+0.24}_{-0.23}$ \mj\ and a radius of $1.091^{+0.028}_{-0.025}$ \rj. Our spectroscopic follow-up also reveals a long-term radial velocity trend for TOI-559, indicating a long-period companion. The statistically significant orbital eccentricity measured for each system suggests that these planets migrated to their current location through dynamical interactions. Interestingly, both planets are also massive ($>$3 \mj), adding to the population of massive giant planets identified by TESS. Prompted by these new detections of high-mass planets, we analyzed the known mass distribution of hot and warm Jupiters but find no significant evidence for multiple populations. TESS should provide a near magnitude-limited sample of transiting hot Jupiters, allowing for future detailed population studies.
\end{abstract}

\section{Introduction}
\label{sec:introduction}

\begin{table*}
\small
\setlength{\tabcolsep}{2pt}
\centering
\caption{Literature and Measured Properties for TOI-558 and TOI-559}
\label{tab:lit}
\begin{tabular}{llccc}
  \hline
  \hline
Other identifiers\dotfill \\
& & TOI-558 & TOI-559 \\
& & TIC 207110080& TIC 209459275 \\
& & TYC 8497-00028-1& TYC 7019-00191-1\\
&TESS Sector & 2,3,29,30 & 4, 31\\
\hline
\hline
Parameter & Description & Value & Value & Source\\
\hline 
$\alpha_{J2000}$\dotfill	&Right Ascension (RA)\dotfill &02:49:09.9601&03:07:16.4958&1\\
$\delta_{J2000}$\dotfill	&Declination (Dec)\dotfill &-58:01:28.9180&-31:09:45.7019&1\\
&                \\
B$_T$\dotfill			&Tycho B$_T$ mag.\dotfill & 12.049 $\pm$ 0.126&11.792$\pm$0.054&2\\
V$_T$\dotfill			&Tycho V$_T$ mag.\dotfill & 11.309 $\pm$ 0.093&11.158$\pm$0.052&2\\
${\rm G}$\dotfill     & Gaia $G$ mag.\dotfill     &11.33$\pm$0.02&10.98$\pm$0.02&1\\
B$_{\rm P}$\dotfill			&Gaia B$_{\rm P}$ mag.\dotfill & 11.58$\pm$0.02&11.30$\pm$0.02&1\\
R$_{\rm P}$\dotfill			&Gaia R$_{\rm P}$ mag.\dotfill & 11.576$\pm$0.02&10.52$\pm$0.02&1\\
${\rm T}$\dotfill     & TESS mag.\dotfill     & 10.988$\pm$0.019& 10.535$\pm$0.018& 3 \\
&                \\
J\dotfill			& 2MASS J mag.\dotfill & 10.581  $\pm$ 0.03	&9.985$\pm$0.023& 4	\\
H\dotfill			& 2MASS H mag.\dotfill & 10.309 $\pm$ 0.03	    &9.719$\pm$0.022& 4	\\
K$_S$\dotfill			& 2MASS ${\rm K_S}$ mag.\dotfill & 10.262 $\pm$ 0.02&9.638$\pm$0.024& 4	\\
&                \\
\textit{WISE1}\dotfill		& \textit{WISE1} mag.\dotfill & 10.216 $\pm$ 0.03 &9.61$\pm$0.03& 5	\\
\textit{WISE2}\dotfill		& \textit{WISE2} mag.\dotfill & 10.248 $\pm$ 0.03 &9.65$\pm$0.03& 5	\\
\textit{WISE3}\dotfill		& \textit{WISE3} mag.\dotfill & 10.236 $\pm$ 0.051&9.61$\pm$0.035& 5	\\
\textit{WISE4}\dotfill		& \textit{WISE4} mag.\dotfill & ---& 9.33$\pm$0.49& 5	\\
&                \\
$\mu_{\alpha}$\dotfill		& Gaia DR2 proper motion\dotfill & 1.071 $\pm$ 0.042 &-23.136$\pm$0.031& 1 \\
                    & \hspace{3pt} in RA (mas yr$^{-1}$)	&&                \\
$\mu_{\delta}$\dotfill		& Gaia DR2 proper motion\dotfill 	&  3.859 $\pm$ 0.042 &-69.698$\pm$0.040& 1 \\
                    & \hspace{3pt} in DEC (mas yr$^{-1}$) &  &                \\
$\pi^\dagger$\dotfill & Gaia Parallax (mas) \dotfill & 2.4850 $\pm$ 0.033$^{\dagger}$ &4.288$\pm$0.037&  1 \\
$v\sin{i_\star}$\dotfill &  Rotational velocity (\kms) \hspace{9pt}\dotfill &7.8$\pm$0.5& 4.08$\pm$0.5  & \S\ref{sec:PFS}\& \ref{sec:CHIRON}               \\
v$_{\rm mac}$\dotfill &  macroturbulent broadening (\kms) \hspace{9pt}\dotfill & 5.9$\pm$0.5 &  4.38$\pm$0.5& \S\ref{sec:PFS}\& \ref{sec:CHIRON}      \\
\\
$U^{*}$\dotfill & Space Velocity (\kms)\dotfill &$3.8\pm 0.1$&$86.0\pm 0.7$&  \S\ref{sec:uvw} \\
$V$\dotfill       & Space Velocity (\kms)\dotfill &$-3.0\pm 0.3$&$-14.5\pm 0.3$&  \S\ref{sec:uvw} \\
$W$\dotfill       & Space Velocity (\kms)\dotfill & $-20.2\pm 0.4$&$4.7\pm 0.5$&  \S\ref{sec:uvw} \\
\hline
\end{tabular}
\begin{flushleft}
 \footnotesize{ \textbf{\textsc{NOTES:}}
 The uncertainties of the photometry have a systematic error floor applied. \\
 $\dagger$ Values have been corrected for the 30 $\mu$as offset as reported by \citet{Lindegren:2018}. \\
  $*$ $U$ is in the direction of the Galactic center. \\
Sources are: $^1$\citet{Gaia:2018},$^2$\citet{Hog:2000},$^3$\citet{Stassun:2018_TIC},$^4$\citet{Cutri:2003}, $^5$\citet{Zacharias:2017}\\
}
\end{flushleft}
\label{tbl:LitProps}
\end{table*}

\par The formation and migration of giant planets in close orbits has been debated extensively. Hot Jupiters (with orbital periods less than 10 days) could theoretically form in a number of ways, with three main formation and migration schemes dominating the literature. It has traditionally been thought that short-period giant planets must form at larger orbital radii and migrate inwards over time \citep{Lin:1996, Rafikov:2006, Dawson:2018}. In order for the formation outcome to be a giant planet, the core needs to form rapidly enough to accrete gas within the lifetime of the proto-planetary disk \citep{Bodenheimer:1986}. Core accretion theories suggest that this atmospheric accretion can only occur in a region of the disk where the core can coalesce enough material to grow to \(\sim\)10 Earth masses -- this critical mass declines as a function of semi-major axis \citep{Piso:2015}. This assumes that the mass of the gaseous envelope becomes greater than the mass of the core \citep{Pollack:1996}. After formation, a giant planet could migrate to a close-in orbit through either gentle migration through the gas disk \citep{Goldreich:1980, Lin:1986, Lin:1996} or more dynamical migration caused by interaction with another planet or star \citep{Rasio:1996, Wu:2003, Fabrycky:2007, Nagasawa:2011, Wu:2011}, after which the planet's orbit could be circularized and shrunk by tidal forces \citep{Naoz:2011, Beauge:2012}. However, more recent models have suggested that hot Jupiters may also form in-situ \citep{Batygin:2016} and show that the period-mass distribution and inner boundary of short-period giant planets could be consistent with predictions for in-situ formation \citep{Bailey:2018}. Other efforts have shown this mass distribution of giant planets to be consistent with high eccentricity migration from dynamical interaction \citep{Matsakos:2016}. The dominance of each of these three formation and migration scenarios remains an open question, and it is likely that a combination of these methods have shaped the hot Jupiter population seen today. Atmospheric characterization is one frontier that may constrain hot Jupiter migration; the measurement of carbon and oxygen abundances in hot Jupiters can be used to trace migration histories \citep{Madhusudhan:2014}.

The discovery of very massive giant planets (> 6 \mj),\footnote{\url{exoplanetarchive.ipac.caltech.edu/}} has raised the question of whether there are meaningful mass boundaries separating giant planets, brown dwarfs, and low-mass stars--specifically, whether there is a particular mass range in which the dominant formation mechanism changes from core accretion to gravitational instability and fragmentation of giant molecular gas clouds. Some studies \citep[e.g.][]{Schlaufman:2018, Moe:2019} have argued that core accretion is the dominant formation mechanism for giant planet companions with masses $M_P<5\ M_J$. Additionally, \citet{Schlaufman:2018} notes that higher host star metallicity is the property associated with core accretion and may indicate that $M_P<5\ M_J$ giant planets may preferentially form via core accretion around metal-rich stars. There also exists a gap in the mass distribution of giant planets very near the threshold of $M_P=7\ M_J$ that \citet{Moe:2019} claim to be a feasible lower mass boundary for disk fragmentation to form an object. \citet{Moe:2019} also highlight that relatively metal-poor host stars seem to preferentially host objects at masses at and above this $M_P=7\ M_J$ threshold. The discovery and characterization of massive giant planets and low-mass brown dwarfs may enable a better understanding of the transition between these formation mechanisms.

\par The observed parameters of a planet and its orbit may be indicative of its formation and migration mechanism. One possible path to determining the dominant mechanism of giant planet migration is to create a complete sample of hot Jupiters with well characterized fundamental parameters (masses and radii, and orbital periods and eccentricities). Statistical population studies of such a sample may provide insight into the dominant evolutionary pathways for giant planets; this type of analysis led to the discovery of the radius valley in small planets \citep{Fulton:2017, Fulton:2018}, supporting the prediction due to photo-evaporation of volatiles \citep{Yelle:2004, Tian:2005, Murray-Clay:2009, Owen:2012, Lopez:2013}.

\par NASA's Transiting Exoplanet Survey Satellite (\tess), launched in 2018, is an all-sky photometric survey with the goal of discovering thousands of new planets around bright, nearby stars \citep{Ricker:2015}. The \tess\ mission has already discovered over a dozen new hot Jupiters, including a few massive systems ($>3$ \mj\ \citealp{Rodriguez:2019a, Nielsen:2020, Rodriguez:2021}), and is expected to be largely complete for giant planets with periods up to 10 days around bright stars \citep{Zhou:2019}. Detailed characterization of new discoveries from \tess\ will help complete the sample of known short-period giant planets, setting the foundation for more robust population studies. 

\par In this paper, we confirm and characterize two short-period giant planets from \tess, TOI-558 b and TOI-559 b. We present the photometric and spectroscopic observations from \tess\ and ground-based facilities in \S\ref{sec:obs}, which we globally model using \texttt{EXOFASTv2} \citep{Eastman:2019} in \S\ref{sec:GlobalModel}.  Further, we examine the existing population of hot Jupiters, studying existing trends in the mass-period distribution and discussing the contribution of \tess\ discoveries (\S\ref{sec:discussion}). Our conclusions are summarized in \S\ref{sec:conclusion}.

\section{Observations and Archival Data}
\label{sec:obs}
{ We confirm and characterize TOI-558 and TOI-559 as planetary systems using \tess\ observations combined with ground-based photometric and spectroscopic follow-up observations from the \tess\ Follow-up Observing Program (TFOP) Working Group.} Table \ref{tab:lit} provides a list of the literature identifiers, magnitudes, and kinematics for TOI-558 and TOI-599.

\begin{figure*}[!ht]
	\centering\vspace{.0in}
	\includegraphics[width=\linewidth]{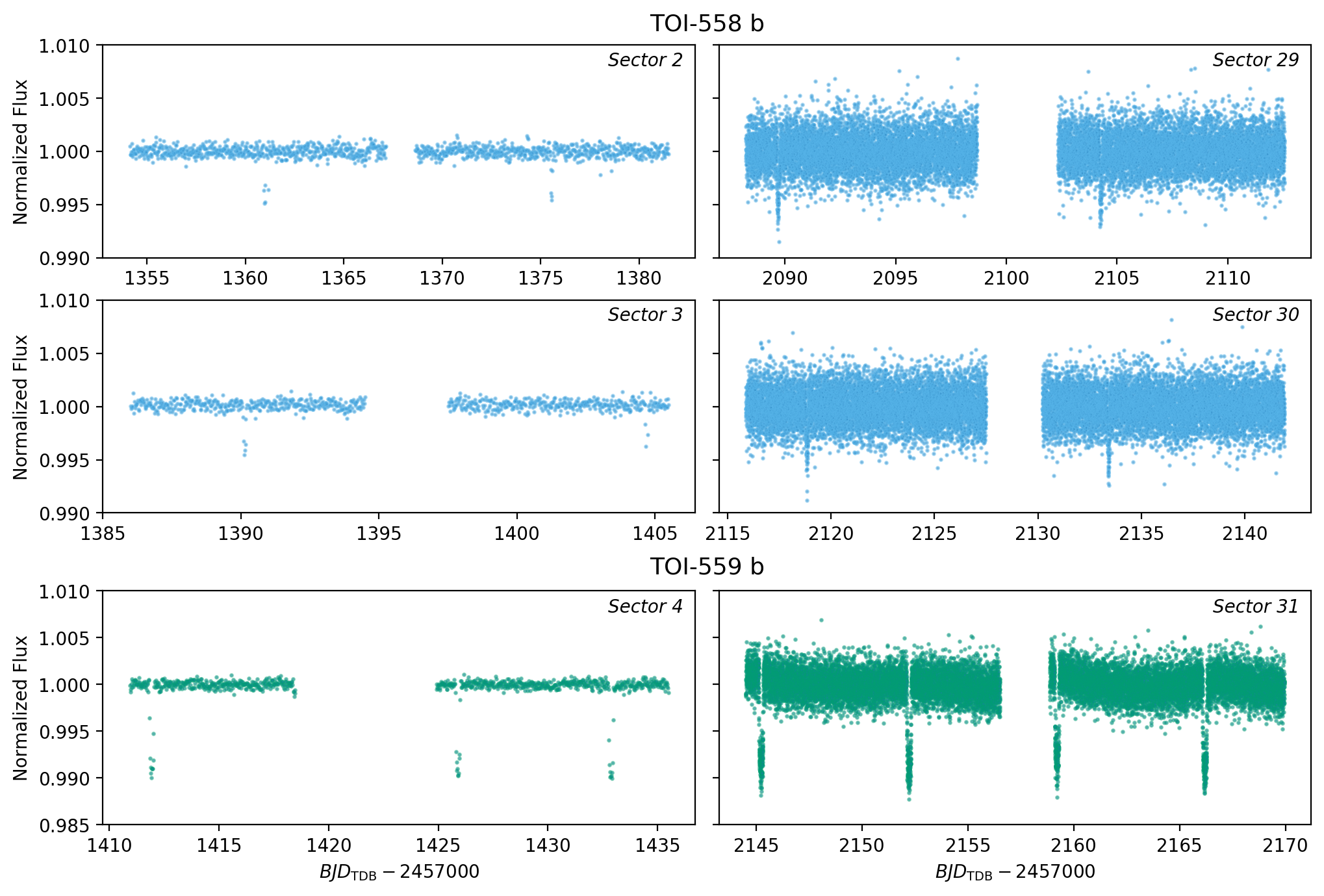}
    \caption{The full corrected light curves from TESS. The discovery light curves (left) extracted from the Full Frame Images are at a 30-minute cadence, from Sectors 2 and 3 for TOI-558 and Sector 4 for TOI-559, { corrected using the quaternions following the description in \S\ref{subsection:tess}.} The additional light curves (right) extracted by the SPOC pipeline are at 2-minute cadence, from Sectors 29-31 of the first extended mission (see \S\ref{subsection:tess}), { corrected with the PDC module. These are not the flattened light curves used for the global fitting.}}
    \label{fig:tess}
\end{figure*}

\begin{figure*}[!ht]
    \centering
    \includegraphics[width=0.5\textwidth]{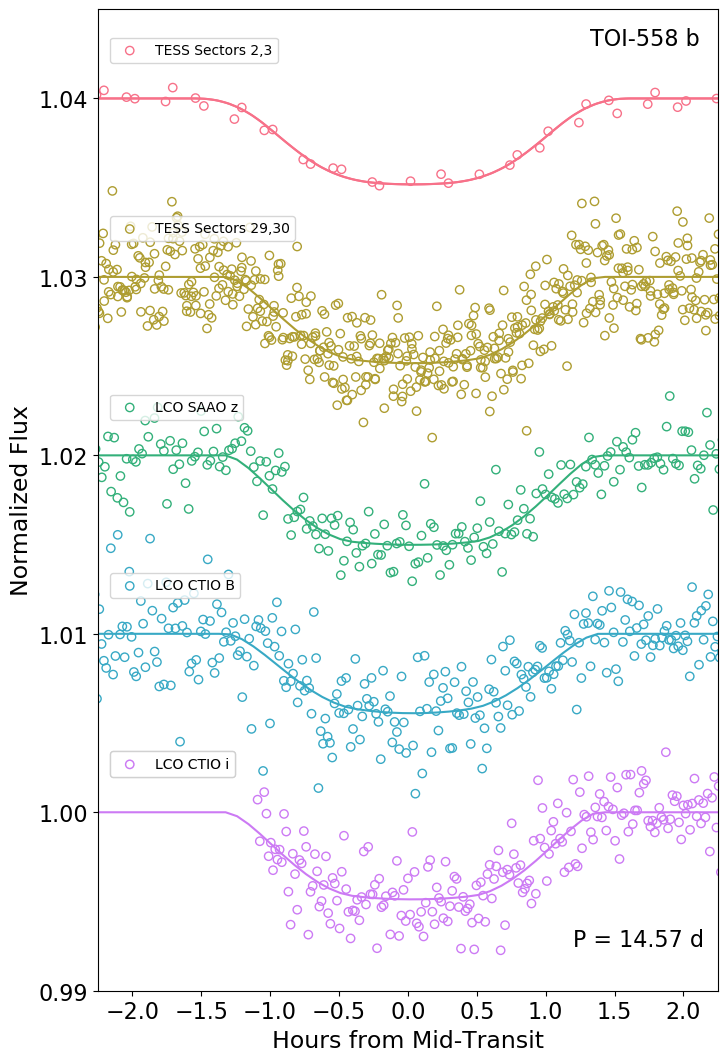}\includegraphics[width=0.481\textwidth]{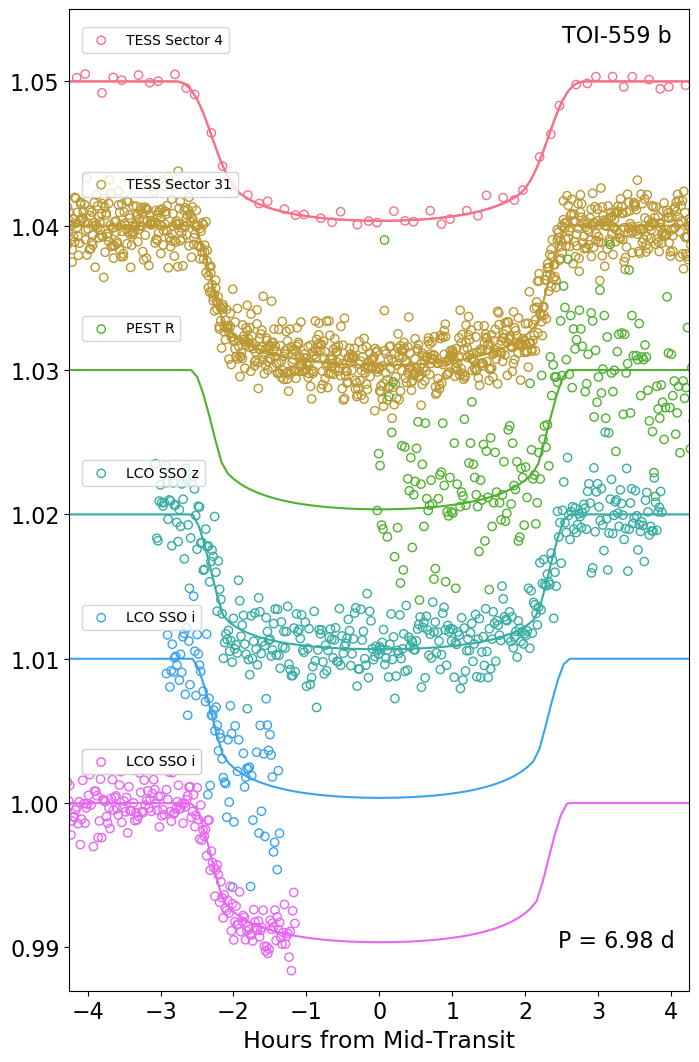}
    \caption{The phase-folded { and detrended} transit light curves for (left) TOI-558 and (right) TOI-559 from \tess\ and the TFOP working group. The solid colored lines correspond to the best-fit model from our global fit (see \S\ref{sec:GlobalModel}).}
    \label{fig:transits}
\end{figure*}

\subsection{\tess\ Photometry}
\label{subsection:tess}
In the two-year primary mission, \tess\ completed 26 observation sectors, each of approximate length $\sim$27 days, covering the southern hemisphere in the first year-long cycle and the northern hemisphere in the second \citep{Ricker:2015}. \tess\ recently began its first extended mission with a similar observation footprint that will cover over 90\% of the sky in total, including a large part of the ecliptic plane. As of UT 2021 January 1, \tess\ has yielded 91 confirmed planets and 2440 planet candidates  { (including planets discovered prior to \tess)}, or \tess\ Objects of Interest (TOIs)\footnote{https://tev.mit.edu/data/collection/193/}.
 
{ \tess\ used four wide-field cameras, each with an f/1.4 aperture, 21 arcsecond pixel scale, and field of view of \(24^{\circ} \times 24^{\circ}\), comprising a total field of view of \(24^{\circ} \times 96^{\circ}\) for each observing sector. \tess\ observations come with a cadence of 20s, 2 minutes, or 30-minute Full Frame Images (FFIs, thought we note the extended mission is now 10 minutes). The 20s and 2 minute cadence targets are preselected before the sector is observed. Unfortunately, neither TOI-558 nor TOI-559 were pre-selected for short-cadence observations during the prime mission (later re-observed in 2 minute cadence during the extended mission; both were observed only in the FFIs, which cover the entire field of view at a 30-minute cadence. }

TOI-558 (TIC 207110080) was observed by Camera 3 in both Sector 2, from UT 2018 August 22 to UT 2018 September 20, and Sector 3, from UT 2018 September 20 to UT 2018 October 18, during \tess's first year of the primary mission. TOI-559 (TIC 209459275) was observed by Camera 2 in Sector 4, from UT 2018 October 18 to UT 2018 November 15. We identified TOI-558 and TOI-559 as planet candidates through a search independent of the \tess\ planet search pipeline, using a standard Box Least Squares algorithm \citep{Kovacs:2002} { and visual examination of candidates from the MIT Quick Look Pipeline (QLP, \citealp{Huang:QLP}}, and both candidates were designated as pre-selected targets for Cycle 3. TOI-558 was then reobserved by \tess\ again during Cycle 3, the first year of the extended mission, in Sector 29 (UT 2020 August 26 to UT 2020 September 22) and Sector 30 (UT 2020 September 22 to UT 2020 October 21) at a cadence of 2 minutes. TOI-559 was reobserved during Sector 31 from UT 2020 October 21 to UT 2020 November 19. The 2-minute observations were inspected by the SPOC team and did not indicate a false positive transit detection \citep{Twicken:2018,Li:2019}.

We extracted and processed light curves from the FFIs using \texttt{Tesscut} and the \texttt{Lightkurve} package for Python \citep{Lightkurve:2018,Brasseur:2019}. The \tess\ Science Processing Operations Center (SPOC) \citep{Jenkins:2016} at NASA Ames Research Center processed the raw FFIs through a pipeline that calibrated the pixels and mapped world coordinate system (WCS) information for each image frame. Our selected apertures included pixels with a mean flux of 80th percentile or greater within a 3-pixel radius of the target's center. We subtracted background scattered light and deblended contamination from nearby stars using a simple target star model. We removed spacecraft systematic effects by decorrelating against the scattered background light and the standard deviation of the quaternion time series following \citet{Vanderburg:2019}. We performed the decorrelation using \texttt{Lightkurve}'s \texttt{RegressionCorrector} utility.  We used the spline-fitting routine {\it Keplerspline}\footnote{\url{https://github.com/avanderburg/keplersplinev2}} \citep{Vanderburg:2014, Shallue:2018} on these light curves to remove any remaining stellar variability, resulting in a flattened light curve. For TOI-559, the baseline fluxes observed during the two orbits of the \tess\ spacecraft in Sector 4 had a significant offset, so we detrended the two orbits separately. We omitted from further consideration all of the data obtained long before or after a transit, leaving roughly one full transit duration prior to each ingress and after each egress (including the full transit). These light curves were then used for the global modeling described in \S \ref{sec:GlobalModel}. 

{ The 2-minute cadence \tess\ light curves for TOI-558 (from Sectors 29 and 30) and TOI-559 (from Sector 31) were extracted by the Science Processing Operations Center (SPOC) pipeline,} based at the NASA Ames Research Center \citep{Jenkins:2016}. Specifically, the data were downloaded, reduced and analyzed by the SPOC pipeline, which included pixel-level calibrations, optimization of photometric aperture, estimation of the total flux contamination from other nearby stars, and extraction of the light curve. To remove systematic effects and instrumental artifacts, the Presearch Data Conditioning (PDC, \citealt{Smith:2012}; \citealt{Stumpe:2014}) module was applied to the extracted SPOC light curve. The resulting processed light curve was run through the SPOC Transiting Planet Search (TPS, \citealt{Jenkins:2002}) to identify any known or additional planet candidates. To remove any remaining low-frequency out-of-transit astrophysical or instrumental variability in the light curves, we use {\it Keplerspline}. We simultaneously fit the spline with a transit model to ensure that the transits were not distorted by the removal of low-frequency variability (see \citealt{Vanderburg:2016b} and \citealt{Pepper:2019})

\begin{table*}[!ht]
\centering
\caption{Ground-based photometry observations from TFOP for TOI-558 and TOI-559 used in the global analysis.}
\label{tab:sg1}
\begin{tabular}{lccccccccccc}
\hline
Date (UT) & Facility & size (m) & Filter & FOV & Pixel Scale &Exp (s) & Additive Detrending\\
\hline
\multicolumn{5}{l}{\textbf{TOI-558}}\\
2019 Sept. 28 & LCO SAAO& 1m & Sloan z$^{\prime}$ & 27$\arcmin$ $\times$27$\arcmin$ & 0.39$\arcsec$&55&airmass, Width T1\\
2019 Oct. 26 & LCO CTIO& 1m & $B$ & 27$\arcmin$ $\times$27$\arcmin$ & 0.39$\arcsec$&34&airmass\\
2020 Nov. 09 & LCO CTIO& 1m & i & 27$\arcmin$ $\times$27$\arcmin$ & 0.39$\arcsec$&25&airmass\\
\multicolumn{5}{l}{\textbf{TOI-559}}\\
2019 Sept. 27 & PEST & 0.3048 & Rc & 27$\arcmin$ $\times$27$\arcmin$ & 1.2$\arcsec$& 60 & None\\ 
2019 Oct. 18 & LCO SSO& 1m & Sloan z$^{\prime}$  & 27$\arcmin$ $\times$27$\arcmin$  & 0.39$\arcsec$& 35 &airmass\\
2020 Aug. 20 & LCO SSO& 1m & Sloan i$^{\prime}$  &27$\arcmin$ $\times$27$\arcmin$  & 0.39$\arcsec$& 25&    none   \\
2020 Aug. 27 & LCO SSO& 1m & Sloan i$^{\prime}$  & 27$\arcmin$ $\times$27$\arcmin$  & 0.39$\arcsec$&25&airmass, sky/pixel T1\\
\hline
\end{tabular}
\begin{flushleft}
  \footnotesize{
    \textbf{\textsc{NOTES:}}
 See \S D in the appendix of \citet{Collins:2017} for a description of each detrending parameter.}
\end{flushleft}
\end{table*}



%

\begin{figure*}[!ht]
	\centering\vspace{.0in}
	\includegraphics[width=0.45\linewidth, trim={2.5cm 13.0cm 9.4cm 8.5cm}, clip]{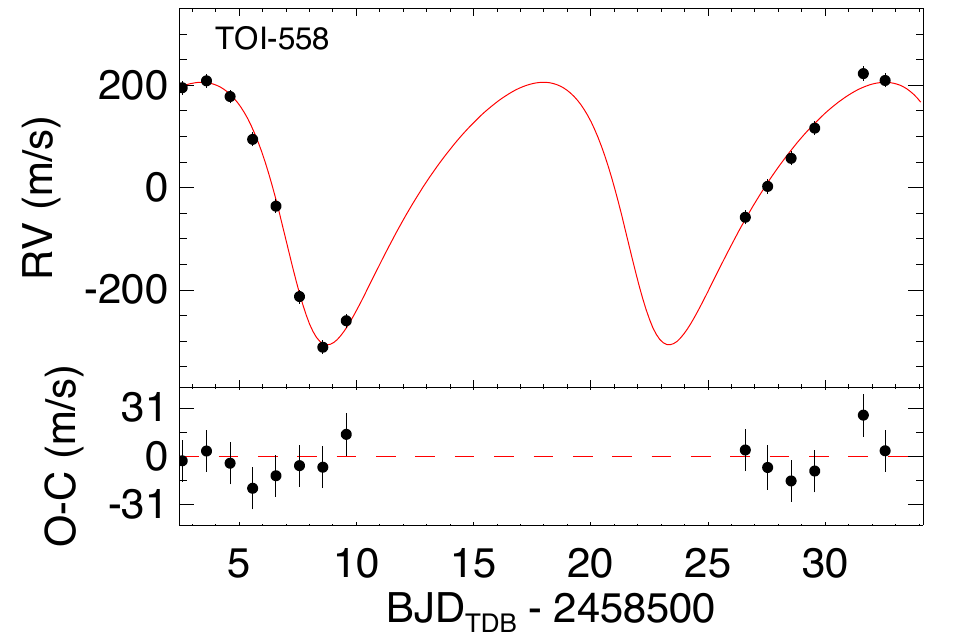}\includegraphics[width=0.45\linewidth, trim={2.5cm 13.0cm 9.4cm 8.5cm}, clip]{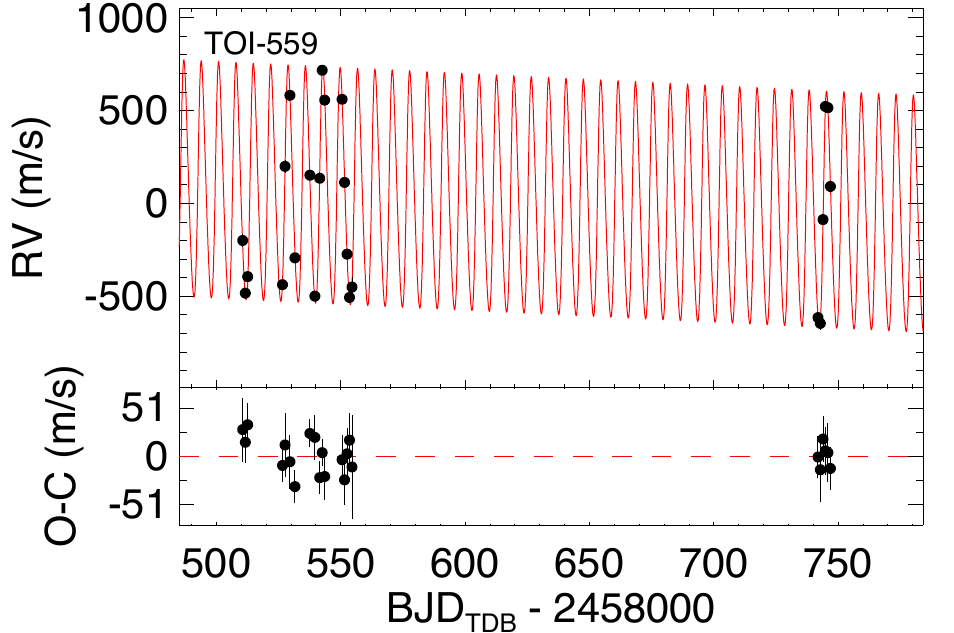}
	\includegraphics[width=0.45\linewidth, trim={2.5cm 13.0cm 9.4cm 8.5cm}, clip]{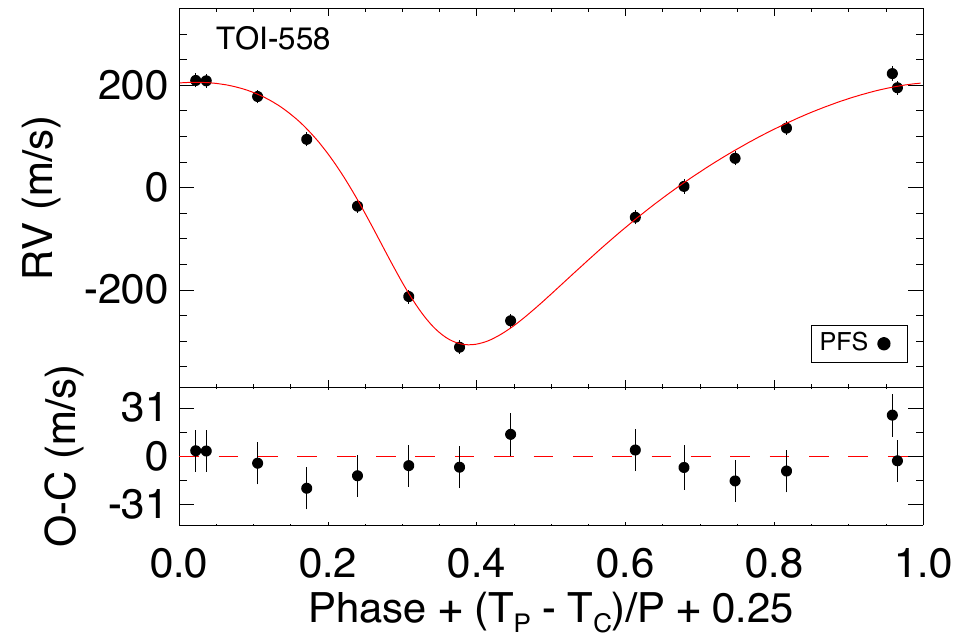}\includegraphics[width=0.45\linewidth, trim={2.5cm 13.0cm 9.4cm 8.5cm}, clip]{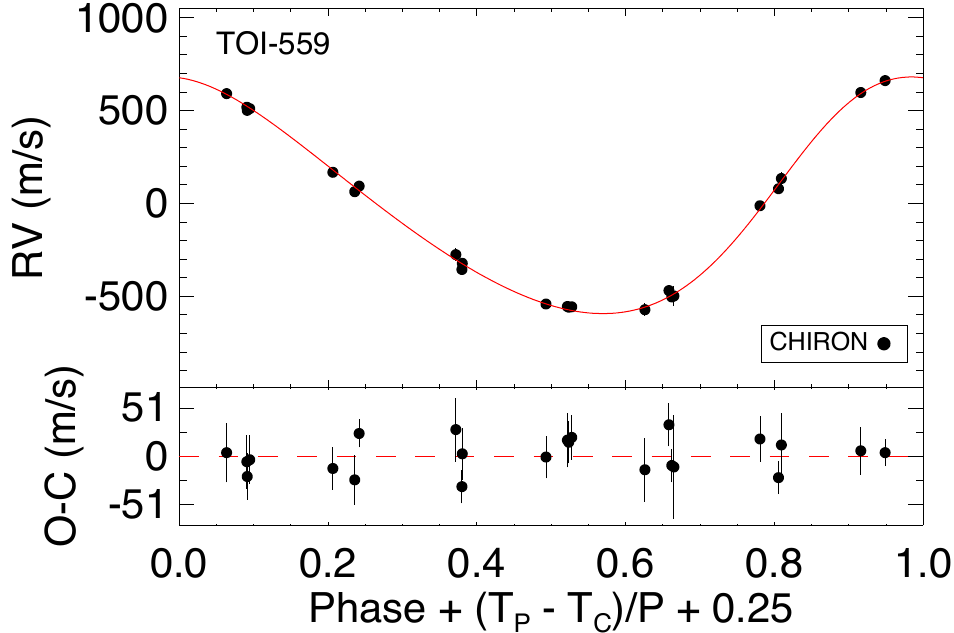}
    \caption{Top) The radial velocity observations over time for (left) \thisstarone\ b from PFS and (right) \thisstartwo\ b from CHIRON. The RVs phase-folded to the best-fit periods are shown above. The EXOFASTv2 model fit is shown in red.}
    \label{fig:RVs}
\end{figure*}

\subsection{Ground-based Photometry from the {\it TESS} Follow-up Observing Program Working Group}
\label{sec:sg1}

To rule out any astrophysical false positives or systematic effects causing the transit events and to refine the timing and transit parameters, we obtained photometric transit follow-up from ground-based telescopes. The \tess\ Follow-up Observing Program (TFOP)\footnote{\url{https://tess.mit.edu/followup/}} Sub Group 1 (SG1), which specializes in ground-based time-series photometry, observed transits of both TOI-558 and TOI-559 with the Las Cumbres Observatory Global Telescope (LCOGT) network of 1-meter telescopes\footnote{\url{https://lco.global/observatory/telescopes/1-m/}} \citep{Brown:2013} and the Perth Exoplanet Survey Telescope (PEST)\footnote{\url{http://pestobservatory.com}}. The observations were scheduled using the \texttt{TAPIR} software package \citep{Jensen:2013}, and all observations but the ones taken by PEST were reduced and lightcurves were extracted using \texttt{AstroImageJ} \citep{Collins:2017}. PEST uses a custom software suite to reduce the images and extract light curves, the \texttt{PEST Pipeline}\footnote{\url{http://pestobservatory.com/the-pest-pipeline/}}. These transit observations and facilities are listed in Table \ref{tab:sg1}. These observations not only extended the baseline, but also provided an independent check on the depth and duration of the transit as compared to what was observed by TESS.

\subsection{PFS Spectroscopy (TOI-558)} \label{sec:PFS}
TOI-558 was observed using the Planet Finder Spectrograph (PFS) on the 6.5-meter Magellan Clay Telescope at Las Campanas Observatory in Chile \citep{Crane:2006,Crane:2008, Crane:2010}, which has been extensively used to follow up and confirm TOIs \citep[e.g.][]{Teske:2020}. We obtained 14 radial velocity (RV) measurements from UT 2019 January 19 to UT 2019 February 18, which are shown in Table \ref{tab:RVs}. { PFS is a high resolution optical (391\ nm to 734\ nm) spectrograph that utilizes an iodine cell to achieve highly precise RV ($<$ 2 m/s) observations. The PFS spectra were reduced and RVs were extracted using a custom IDL pipeline  \citep{Butler:1996}. These observations were taken with the new 10k$\times$10k CCD, with a 0.3$\arcsec$ slit, and at 3$\times$3 binning with a resolving power of (R$\sim$110000). While PFS can achieve sub-1 \ms\ precision, we chose shorter exposures with a typical RV precision of $\sim$5 \ms\ since our targets have very large RV semi-amplitudes ($>$30 \ms). }


We derived stellar parameters, specifically the host star's metallicity, for TOI-558 from the iodine free template spectrum obtained with PFS. The spectrum, in the region of 5000 - 5500 \text{\AA}, was analyzed with the ZASPE package \citep{Brahm:2017}, which performs a model comparison between the observed spectrum and a grid of the PHOENIX stellar atmospheres synthetic spectra \citep{Husser:2013}. ZASPE weights spectral regions based on their pre-determined importance to the stellar parameter determination, and varies the depths of those spectral regions with a Monte Carlo analysis to determine the uncertainties and covariance of the derived stellar parameters. The resulting best fit metallicity was \feh\ = -0.020$\pm$0.066 dex, which we use as a prior on the global fit (see \S\ref{sec:GlobalModel}). Using the PFS template, we measured the $v\sin{i_\star}$ and macroturbulent broadening for TOI-558 following the methodology in \citet{Zhou:2018}. We measured $v\sin{i_\star}$ for TOI-558 to be be 4.1$\pm$0.5 \kms\ and v$_{\rm mac}$ to be 4.4$\pm$0.5 \kms.

\subsection{TRES Spectroscopy (TOI-559)} \label{sec:TRES}
Reconnaissance spectroscopic follow-up observations of TOI-559 were taken on three separate epochs at a resolving power of R$\sim$44,000 using the 1.5m Tillinghast Reflector Echelle Spectrograph \citep[TRES;][]{furesz:2008}\footnote{\url{http://www.sao.arizona.edu/html/FLWO/60/TRES/GABORthesis.pdf}}. TRES is located at the the Fred L. Whipple Observatory (FLWO) on Mt. Hopkins, AZ. The reduction and RV extraction pipeline details are described in \citet{Buchhave:2010} and \citet{Quinn:2012}. With only three observations, we do not include these RVs in the global analysis (see \S\ref{sec:GlobalModel}). Nevertheless the extracted RVs yielded a semi-amplitude consistent with the global analysis. The TRES spectra were also used as an independent check on the metallicity from CHIRON. Using the Stellar Parameter Classification (SPC) package \citep{Buchhave:2012}, we derive a metallicity for TOI-559 of \feh\ = -0.24$\pm$0.08\ dex, consistent with what was found using CHIRON. Additionally, the TRES absolute velocity for TOI-559 was -14.03$\pm$0.42\ \kms, consistent with the Gaia DR2 results.

\begin{table}[!ht]
\centering
\caption{Radial Velocities for TOI-558 and TOI-559}
\begin{tabular}{ccccc}
\hline
Target &\(BJD_{TDB} (days)\) & \(RV (m s^{-1})\) & \(\sigma_{RV} (m s^{-1})\)& Facility\\
\hline
TOI-558 &2458502.57047 & 137.8 & 4.6 & PFS\\
TOI-558 &2458503.60715 & 151.3 & 4.4 & PFS\\
TOI-558 &2458504.61284 & 120.7 & 5.0 & PFS\\
TOI-558 &2458505.57474 & 37.0 & 5.2 & PFS\\
TOI-558 &2458506.56937 & -93.4 & 4.6 & PFS\\
TOI-558 &2458507.57311 & -270.0 & 5.0 & PFS\\
TOI-558 &2458508.56806 & -368.9 & 4.7 & PFS\\
TOI-558 &2458509.56851 & -317.2 & 6.2 & PFS\\
TOI-558 &2458526.58483 & -115.2 & 5.1 & PFS\\
TOI-558 &2458527.53828 & -54.8 & 6.8 & PFS\\
TOI-558 &2458528.53871 & 0.0 & 4.8 & PFS\\
TOI-558 &2458529.54633 & 59.0 & 5.4 & PFS\\
TOI-558 &2458531.61868 & 165.5 & 5.7 & PFS\\
TOI-558 &2458532.54853 & 152.1 & 5.2 & PFS\\
\hline
TOI-559 & 2458510.54861 & -15924.9 & 33.2 & CHIRON\\
TOI-559 & 2458511.60839 & -16208.5 & 21.6 & CHIRON\\
TOI-559 & 2458512.54877 & -16119.7 & 22.1 & CHIRON\\
TOI-559 & 2458526.54291 & -16162.9 & 17.0 & CHIRON\\
TOI-559 & 2458527.57189 & -15526.1 & 33.6 & CHIRON\\
TOI-559 & 2458529.53872 & -15143.5 & 28.0 & CHIRON\\
TOI-559 & 2458531.55750 & -16018.5 & 16.7 & CHIRON\\
TOI-559 & 2458537.57692 & -15573.6 & 13.7 & CHIRON\\
TOI-559 & 2458539.57232 & -16224.7 & 22.9 & CHIRON\\
TOI-559 & 2458541.51279 & -15590.0 & 16.4 & CHIRON\\
TOI-559 & 2458542.51358 & -15008.6 & 13.2 & CHIRON\\
TOI-559 & 2458543.51021 & -15169.4 & 23.8 & CHIRON\\
TOI-559 & 2458550.51777 & -15165.0 & 25.0 & CHIRON\\
TOI-559 & 2458551.50301 & -15612.5 & 25.4 & CHIRON\\
TOI-559 & 2458552.51271 & -15998.8 & 26.6 & CHIRON\\
TOI-559 & 2458553.49918 & -16231.8 & 28.1 & CHIRON\\
TOI-559 & 2458554.49890 & -16175.8 & 54.2 & CHIRON\\
TOI-559 & 2458741.86403 & -16340.3 & 21.7 & CHIRON\\
TOI-559 & 2458742.79259 & -16371.4 & 33.3 & CHIRON\\
TOI-559 & 2458743.87250 & -15812.6 & 23.1 & CHIRON\\
TOI-559 & 2458744.81891 & -15203.6 & 24.5 & CHIRON\\
TOI-559 & 2458745.85030 & -15210.3 & 30.6 & CHIRON\\
TOI-559 & 2458746.84768 & -15634.3 & 21.6 & CHIRON\\
\hline
TOI-559 & 2458511.60506 & -896.80 & 20.02 & TRES\\
TOI-559 & 2458515.62512 & 218.75 & 20.29 & TRES\\
TOI-559 & 2458738.98177 & 104.78 & 34.18 & TRES\\       
\hline
\end{tabular}
\label{tab:RVs}
\begin{flushleft}
 \footnotesize{ \textbf{\textsc{NOTES:}}
 The median absolute RV has been subtracted off the PFS and TRES RVs. 
 }
\end{flushleft}
\end{table}

\subsection{CHIRON Spectroscopy (TOI-559)} \label{sec:CHIRON}
TOI-559 was observed with the CTIO High Resolution spectrometer (CHIRON) on the CTIO 1.5-meter telescope \citep{Tokovinin:2013}. CHIRON covers a wavelength range of 420\ nm to 880\ nm, with a resolving power of R$\sim$80,000. The RV measurements were extracted from the CHIRON spectra using a least-squares deconvolution technique described in \citet{Donati:1997, Zhou:2020}; the 22 RVs were taken between UT 2019 January 27 and UT 2019 September 20 and are shown in Table \ref{tab:RVs}. We check that the line broadening velocity is not correlated with the measured radial velocities. We also note that both TOI 558 and 559 are slowly rotating stars, with rotational broadening velocities of 8 \kms\ and 4 \kms\ respectively. For stellar activity to affect our velocities at the 200m/s level, as is our detected Doppler orbit, the stars should exhibit significant photometric modulation at the $>$2\% level. We do not see any large stellar activity signatures in the TESS light curves, consistent with our interpretation that these target stars are quiet at the level suitable for our detections of their Doppler orbits.



We also use the CHIRON spectra to determine some constraints on the host star's metallicity and $v\sin{i_\star}$. The spectra were matched against an interpolated grid of $\sim10,0000$ observed spectra from the TRES database, previously classified using the Spectral Classification Pipeline \citep{Buchhave:2012}. This library is interpolated using a gradient boost classifier algorithm in the \emph{scikit-learn} machine learning package. The CHIRON observed spectrum is then convolved against a Gaussian profile such that it matches the spectral resolution of observations in this library ($R=44,000$). We measure the metallicity of TOI-559 to be \feh\ = -0.22$\pm$0.11 dex, the effective temperature to be \teff\ = 5784 $\pm$ 50\ K, and the surface gravity to be \logg\ = 4.18 $\pm$ 0.10\ (cgs). We only use the metallicity as a Gaussian prior in the global fit, allowing the fit to constrain the host star's effective temperature and surface gravity using the spectral energy distribution and transit shape, respectively (see \S\ref{sec:GlobalModel}). We also derive the $v\sin{i_\star}$ for TOI-559 to be 7.8$\pm$0.5\ \kms\ and v$_{\rm mac}$ to be 5.9$\pm$0.5\ \kms\ following \citet{Zhou:2018}.

\begin{figure*}[!ht]
    \centering
    \includegraphics[width=0.45\textwidth]{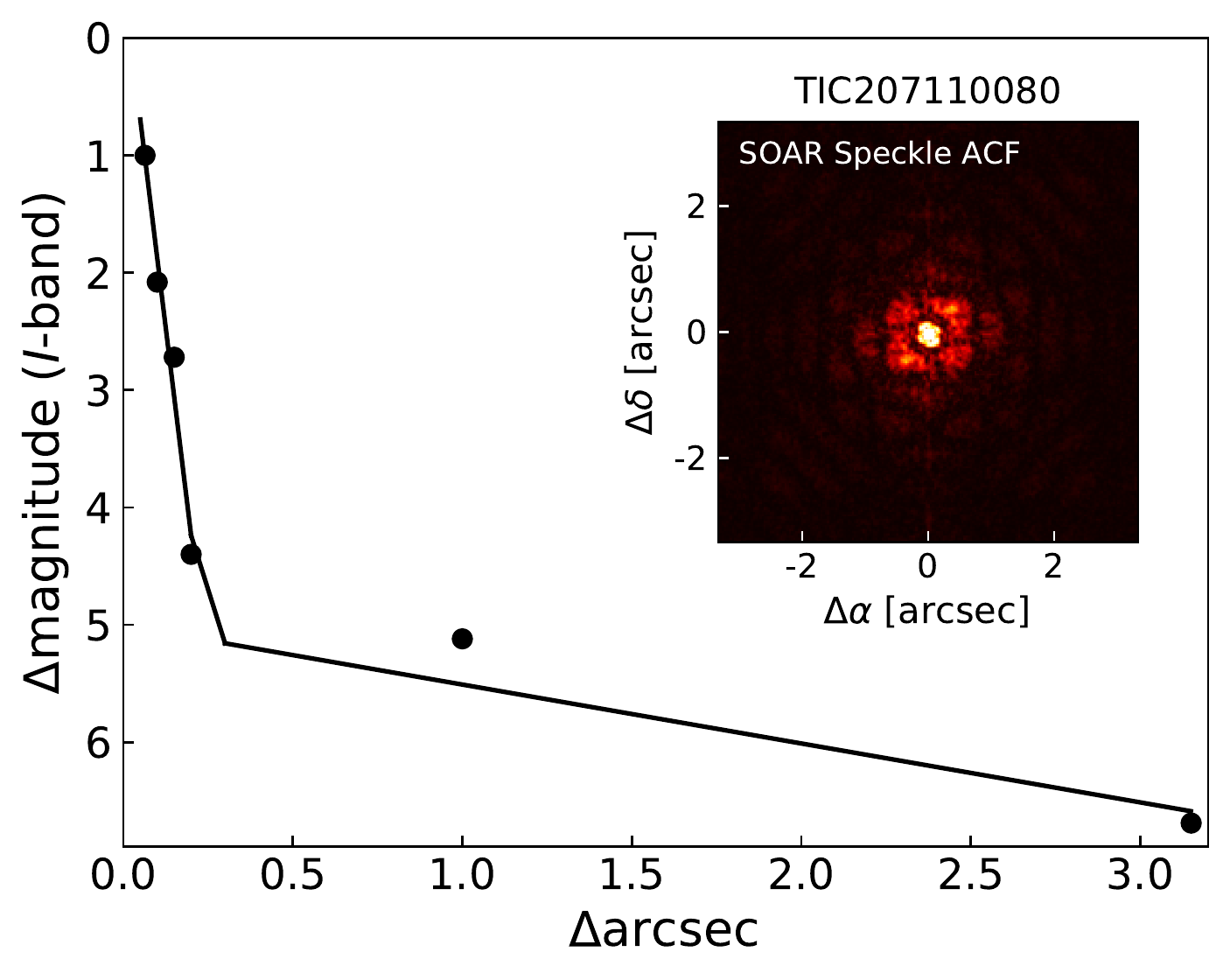}\includegraphics[width=0.45\textwidth]{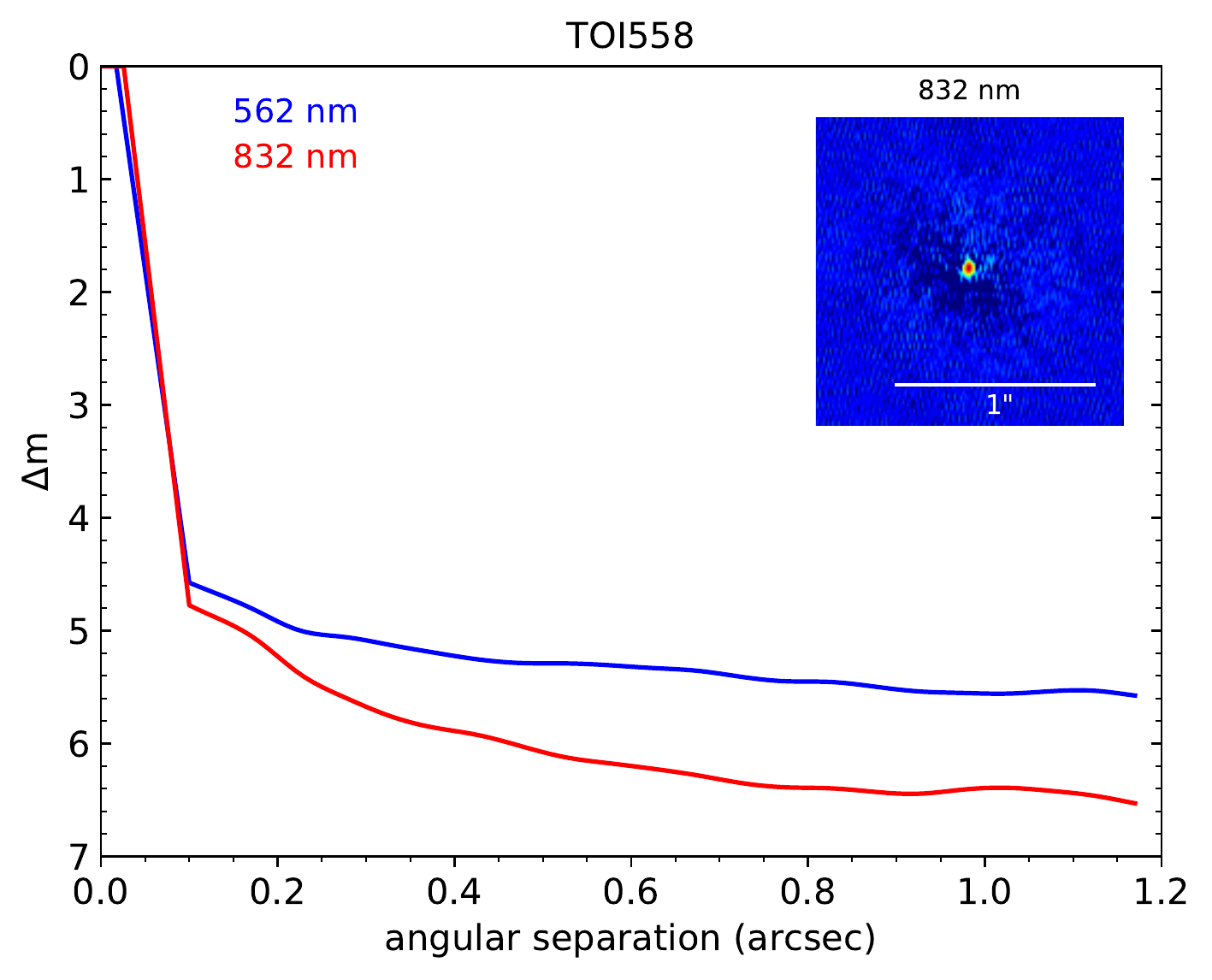}\\
    \includegraphics[width=0.45\textwidth]{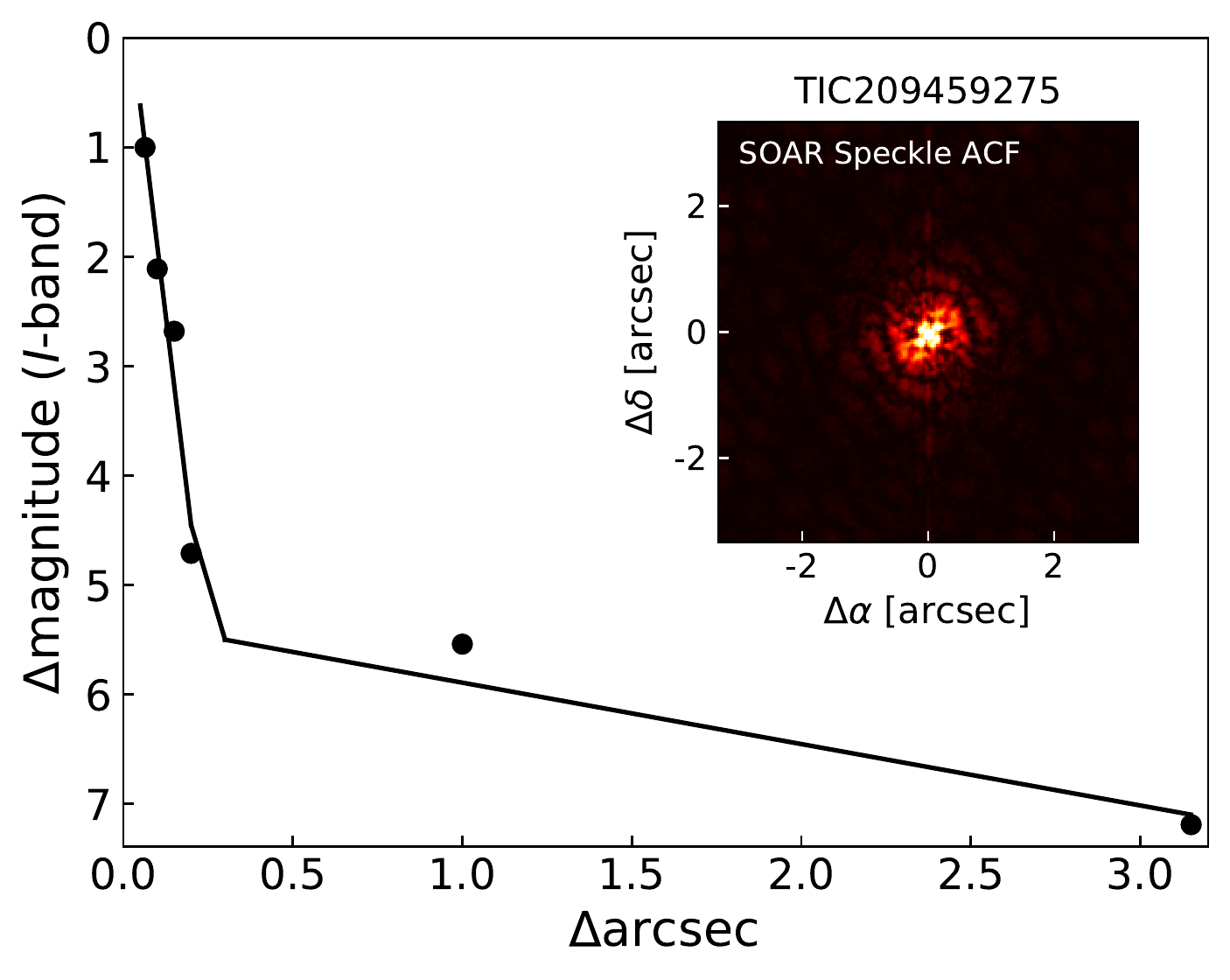}\includegraphics[width=0.45\textwidth]{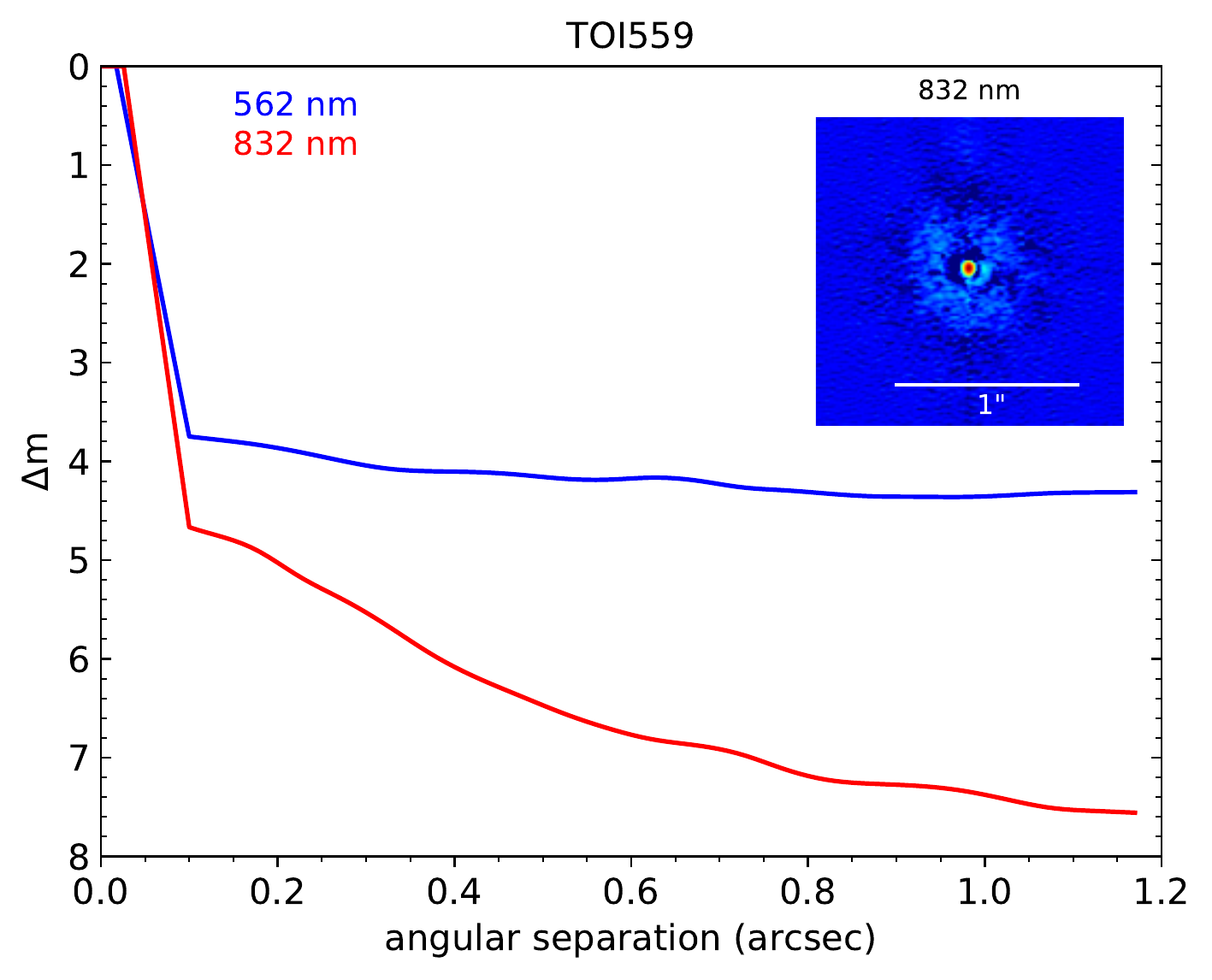}\\
    
    \caption{The (left) Speckle interferometric observations for TOI-558 and TOI-559 of the two targets from the Southern Astrophysical Research Telescope (SOAR). The autocorrelation function is shown inset the contrast curve from SOAR. { The (right) Gemini South Zorro and Gemini North `Alopeke speckle imaging 5-sigma contrast curves are shown along with the reconstructed images (embedded) of TOI-558 and TOI-559.}}
    \label{fig:SOAR}
\end{figure*}

\subsection{High Resolution Speckle Imaging}
\label{sec:SOAR}

It is difficult to rule out the possibility of blended companion stars using TESS data alone given the size of the pixels. Contamination from blended stars can cause a false positive transit signal on the planetary candidate host star or affect the derived planetary radius \citep{Ciardi:2015,Ziegler:2018}. { To check for very nearby stars not resolved by seeing-limited images and Gaia and account for any blends that would be included in the spectra, we obtained high-resolution speckle imaging of TOI-558 and TOI-559 from the Southern Astrophysical Research Telescope (SOAR) \citep{Tokovinin:2018}.} TOI-558 and TOI-559 were both observed on UT 2019 February 18, and had a sensitivity of $\Delta$Mag = 6.7 and 7.2 at 1$\arcsec$, respectively.  Figure \ref{fig:SOAR} displays the reconstructed images as well as the limiting magnitude difference versus on-sky distance from the center of the target star. We see no signs of any nearby close (within 3 arcseconds) companions in the SOAR observations of TOI-558 or TOI-559. For a detailed description of the observing strategy for \tess\ targets see \citet{Ziegler:2020}.

{ Using the Zorro instrument mounted on the 8-meter Gemini South telescope on Cerro Pachon in Chile, we observed TOI-558 on UT 2020 December 23 and 29. The first observation had poor seeing, so we show the December 29th observation in Figure \ref{fig:SOAR}.} TOI-559 was observed using the `Alopeke instrument on UT 2019 October 09. `Alopeke simultaneously observes in blue ($\frac{\lambda}{\Delta\lambda}$ = 562/54\ nm) and red ($\frac{\lambda}{\Delta\lambda}$ = 832/40\ nm) band passes, with inner working angles of 0.026$\arcsec$ for the blue and 0.017$\arcsec$ for the red. The instrument has a pixel scale of 0.01$\arcsec$. Three thousand 0.06-second images were obtained and combined for each star, and the Fourier analysis described in \citet{Howell:2011} was performed on the combined image. The `Alopeke observations confirm and extend to smaller inner working angles the results seen by SOAR, in that TOI-559 is a single star with no signs of any previously unknown companions to within the 5-sigma contrast limits obtained (see Figure \ref{fig:SOAR}). The observations had a sensitivity of $\Delta$mag = 5.557 for the blue and 7.375 for the red, at 1$\arcsec$ for TOI-559 and $\Delta$mag = 4.355 and 6.394 for TOI-558. { The observations and full contrast curves can be found on \url{https://exofop.ipac.caltech.edu/tess/}.}

\subsection{Galactic Locations, Kinematics, Orbits, and Populations}
\label{sec:uvw}

We used the parallaxes, proper motions, radial velocities, and associated uncertainties of TOI-558 and TOI-559 from the Gaia DR2 catalog \citep{Gaia:2018} to determine the location, kinematics, orbits, and associations of each system with known stellar populations following the analysis methodology performed by \citet{Burt:2020}. We corrected the DR2 parallaxes and uncertainties following \citet{Lindegren:2018}. We then used these parallaxes to estimate the distances to the systems. These distances and their uncertainties where then used in combination with the DR2 proper motions and radial velocities to determine the heliocentric UVW velocities of the host stars.  We determined the UVW velocities with respect to the Local Standard of Rest (LSR) using the determination of the sun's motion relative to the LSR by \citet{Coskunoglu:2011}. We adopt a coordinate system such that positive $U$ is toward the Galactic center. These UVW values are shown in Table 1.
 
For each system, we estimated its $Z$ height relative to the sun, and then corrected for the $Z_\odot \simeq 30~{\rm pc}$ offset of the sun from the Galactic plane as determined by \citet{Bovy:2017} based on local giants.  We use the UVW velocities (with respect to the LSR) to estimate the likelihood that the star belongs to thin disk, thick disk, halo, or Hercules stream, using the categorization criteria of \citet{Bensby:2014}. We use the Galactic orbits estimated by \citet{Mackereth:2018}, and report estimates of the orbital parameters (apogalacticon, perigalaciton, eccentricity, and maximum excursion perpendicular to the plane).  We estimated the spectral type of each host star using their effective temperatures (as given in Table 4) and the relations of \citet{Pecaut:2013}. We then compared the position and orbits of the two systems to the scale height $h_Z$ of stars of similar spectral type as determined by \citet{Bovy:2017}.
 
We also considered whether either of the systems belong to any of the known nearby young associations using the BANYAN $\Sigma$ (Bayesian Analysis for Nearby Young AssociatioNs $\Sigma$) tool \citep{Gagne:2018}. The BANYAN $\Sigma$ estimator assigned both hosts to be `field' stars.

TOI-558 is at a distance of $d=402\pm5~{\rm pc}$ from the sun, consistent with the posterior value listed in Table 4.  Its vertical distance from the Galactic plane is $Z+Z_\odot \simeq -291~{\rm pc}$.  It has Galactic velocities with respect to the LSR of $(U,V,W)=(3.8\pm 0.1,-3.0\pm 0.3, -20.2\pm 0.4)~{\rm km~s^{-1}}$. According to the categorization of \citet{Bensby:2014}, the system has a $\sim 98\%$ probability of belonging to the thin disk.  The Galactic orbit has a perigalacticon of $R_p=7.04~{\rm kpc}$, and apogalacticon of $R_a=8.07~{\rm kpc}$, an eccentricity of $e=0.07$, and a maximum $Z$ excursion from the Galactic plane of $Z_{\rm max}=460~{\rm pc}$. Thus, the orbit is consistent with the current location of the system. The scale height of stars of similar spectral type (F5.5V) is only $85~{\rm pc}$. Nevertheless, there is a non-negligible probability that a star belonging to this population can have a maximum excursion above the plane that is several scale heights. 

TOI-559 is at distance of $d=233\pm2~{\rm pc}$ from the sun, consistent with the posterior value listed in Table \ref{tab:exofast}.  Its vertical distance from the Galactic plane is $Z+Z_\odot \simeq -172~{\rm pc}$.  It has Galactic velocities with respect to the LSR of $(U,V,W)=(86.0\pm 0.7,-14.5\pm 0.3, 4.7\pm 0.5)~{\rm km~s^{-1}}$. According to the categorization of \citet{Bensby:2014}, the system has a $\sim 92\%$ probability of belonging to the thin disk, and an $\sim 8\%$ probability of belonging to the thick disk.  The Galactic orbit has a perigalacticon of $R_p=5.32~{\rm kpc}$, and apogalacticon of $R_a=10.3~{\rm kpc}$, an eccentricity of $e=0.022$, and a maximum $Z$ excursion from the Galactic plane of $Z_{\rm max}=320~{\rm pc}$. Thus the orbit is consistent with the current location of the system and with the scale height of $108~{\rm pc}$ for stars of similar spectral type (G0V). Although this system has a non-negligible probability of belonging to the thick disk, it is nevertheless more likely to be a member of the thin disk.  We estimate an age of $\sim 7$~Gyr for this system { from our global analysis Table \ref{tab:exofast}}, which may explain its relatively large value of the eccentricity and maximum vertical excursion above the plane of the orbit.

\section{EXOFAST\lowercase{v}2 Global Fit for \thisstarone and \thisstartwo} 
\label{sec:GlobalModel}
In order to characterize the planetary systems, we modeled the observations obtained in \S\ref{sec:obs} with \texttt{EXOFASTv2}, a global fitting suite for exoplanets \citep{Eastman:2013, Eastman:2019} to simultaneously fit the \tess\ and TFOP SG1 photometry and the PFS and CHIRON RVs. \texttt{EXOFASTv2} uses a differential evolution Markov Chain Monte Carlo (MCMC) to simultaneously model the star and planet globally and self-consistently. For our fits of TOI-558 and TOI-559, we conducted a fit of the Spectral Energy Distribution (SED) of the host star (see Table \ref{tab:lit} for a list of the broadband photometric measurements used in the SED analysis) simultaneously with the available radial velocities and photometry. We imposed Gaussian priors on the \textit{Gaia} parallaxes \citep{Gaia:2018} (accounting for the 30$\mu$ offset as reported by \citealp{Lindegren:2018}) and the stellar metallicities obtained from spectroscopy (\feh\ = -0.020$\pm$0.066 for TOI-558, $-0.22\pm0.11$ for TOI-559, see \S\ref{sec:CHIRON} \& \ref{sec:PFS}), and an upper limit on the maximum line of sight extinction (0.06169 and 0.04154) according to \citet{Schlegel:1998, Schlafly:2011}. With the addition of catalogued broadband photometry (see Table \ref{tab:lit}) and SED model constraints, the fit provides a precise constraint on the stellar radius ($R_\star$). Within the fit, \texttt{EXOFASTv2} placed a lower bound on the precision ($\sim2\%$) of the bolometric flux (F$_{\rm bol}$) for the SED, which corresponds to the variations in F$_{\rm bol}$ from different calculation techniques \citep{Zinn:2019}. \texttt{EXOFASTv2} uses the MESA Isochrones and Stellar Tracks (MIST) stellar evolution models \citep{Paxton:2011,Paxton:2013,Paxton:2015,Choi:2016,Dotter:2016}, thereby encoding the physics of stellar evolution, where the global model is penalized for large differences from MIST-predicted stellar values. We ran MCMC fits for both systems, with strict convergence criteria of a Gelman Rubin statistic of less than 1.01 and at least 1000 independent draws in each parameter. We also fit for a dilution term on the \tess\ observations. Specifically, we adopt a Gaussian prior on the contamination ratio equal to that reported by the TESS Input Catalog (TIC, \citealp{Stassun:2018_TIC}), with a dispersion of 10\%. This assumes that the \tess\ light curves have been corrected for known companions in the aperture to better than 10\%. Although we deblend the FFI light curve, the SPOC pipeline corrects the 2-minute light curve, and no unknown companions were detected in our high-resolution imaging (see \S\ref{sec:SOAR}). This provides an independent check on those corrections and properly propagate uncertainties. In both cases, the fitted dilution found by \texttt{EXOFASTv2} is consistent with zero. The TFOP SG1 photometry for each system was detrended within the full fit using an additive model and the detrending parameters seen in Table \ref{tbl:LitProps}. See \citet{Collins:2017} appendix D for a description of each detrending parameter listed. The fitted transit data for TOI-558 and TOI-559 are shown in Figure \ref{fig:transits}, the RV fit is shown in Figure \ref{fig:RVs}, and the resulting median values and 1-sigma uncertainties for all fitted stellar and planetary parameters are displayed in Tables \ref{tab:exofast} and \ref{tab:exofast_other}. At the top of Table \ref{tab:exofast} is a list of the priors used in the fit. See \citet{Eastman:2019} for a full list of the fitted and derived parameters from \texttt{EXOFASTv2} and any bounds on fitted parameters. 

\begin{table*}
\scriptsize
\centering
\caption{Median values and 68\% confidence interval of the posterior distribution for the global models}
\begin{tabular}{llccccccc}
  \hline
  \hline
\multicolumn{2}{l}{Priors:}& TOI-558 b &TOI-559 b \\
\hline\\
Gaussian & $\pi$ Gaia Parallax (mas) \dotfill & 2.53691$\pm$0.04045 & 4.28820$\pm$0.03673\\
Gaussian & $[{\rm Fe/H}]$ Metallicity (dex) & -0.02$\pm$0.07 & -0.22$\pm$0.11 \\
Upper Limit & $A_V$ V-band extinction (mag) & 0.0617 & 0.0415 \\
Gaussian$^{\prime}$& $D_T$ Dilution in {\it Tess}  \dotfill &0.00000$\pm$0.000317&0.00000$\pm$0.000102\\
  \hline
  \hline
  Parameter & Units & Values & & & \\
~~~~$M_*$\dotfill &Mass (\msun)\dotfill & $1.349^{+0.064}_{-0.065}$& $1.026\pm0.057$\\
~~~~$R_*$\dotfill &Radius (\rsun)\dotfill & $1.496^{+0.042}_{-0.040}$& $1.233^{+0.028}_{-0.026}$\\
~~~~$L_*$\dotfill &Luminosity (\lsun)\dotfill & $3.52^{+0.16}_{-0.14}$& $1.688^{+0.087}_{-0.069}$\\
~~~~$F_{Bol}$\dotfill &Bolometric Flux$\times$10$^{-10}$ (cgs)\dotfill & $6.99^{+0.26}_{-0.22}$& $9.92^{+0.49}_{-0.37}$\\
~~~~$\rho_*$\dotfill &Density (g cm$^{-3}$)\dotfill & $0.568^{+0.054}_{-0.051}$& $0.774^{+0.053}_{-0.058}$\\
~~~$\log{g}$\dotfill &Surface gravity (cgs)\dotfill & $4.218^{+0.030}_{-0.031}$& $4.268^{+0.024}_{-0.028}$\\
~~~~$T_{\rm eff}$\dotfill &Effective Temperature (K)\dotfill & $6466^{+95}_{-93}$& $5925^{+85}_{-76}$\\
~~~~$[{\rm Fe/H}]$\dotfill &Metallicity (dex)\dotfill & $-0.004^{+0.059}_{-0.055}$& $-0.069^{+0.065}_{-0.079}$\\
~~~~$[{\rm Fe/H}]_{0}$\dotfill &Initial Metallicity \dotfill & $0.137^{+0.051}_{-0.049}$& $-0.001^{+0.063}_{-0.068}$\\
~~~$Age$\dotfill &Age (Gyr)\dotfill & $1.79^{+0.91}_{-0.73}$& $6.8^{+2.5}_{-2.0}$\\
~~~~$EEP^\ddagger$\dotfill &Equal Evolutionary Phase \dotfill & $345^{+22}_{-14}$& $414^{+14}_{-19}$\\
~~~~$A_V$\dotfill &V-band extinction (mag)\dotfill & $0.033^{+0.020}_{-0.022}$& $0.023^{+0.013}_{-0.015}$\\
~~~~$\sigma_{SED}$\dotfill &SED photometry error scaling & $1.02^{+0.34}_{-0.22}$& $0.99^{+0.43}_{-0.25}$\\
~~~~$\varpi$\dotfill &Parallax (mas)\dotfill & $2.491\pm0.032$& $4.289^{+0.036}_{-0.037}$\\
~~~~$d$\dotfill &Distance (pc)\dotfill & $401.4^{+5.3}_{-5.1}$& $233.2\pm2.0$\\
~~~~$\dot{\gamma}$\dotfill &RV slope (m/s/day)\dotfill & ---& $-0.650^{+0.064}_{-0.065}$\\
\multicolumn{2}{l}{Planetary Parameters:}&&\\
~~~~$P$\dotfill &Period (days)\dotfill & $14.574071\pm0.000026$& $6.9839095\pm0.0000051$\\
~~~~$R_P$\dotfill &Radius (\rj)\dotfill & $1.086^{+0.041}_{-0.038}$& $1.091^{+0.028}_{-0.025}$\\
~~~$M_P$\dotfill &Mass (\mj)\dotfill & $3.61\pm0.15$& $6.01^{+0.24}_{-0.23}$\\
~~~~$T_0^\star$\dotfill &Optimal conjunction Time (\bjdtdb)\dotfill & $2458871.07253\pm0.00053$& $2458893.81305\pm0.00023$\\
~~~~$a$\dotfill &Semi-major axis (AU)\dotfill & $0.1291^{+0.0020}_{-0.0021}$& $0.0723\pm0.0013$\\
~~~~$i$\dotfill &Inclination (Degrees)\dotfill & $86.24^{+0.19}_{-0.22}$& $89.08^{+0.52}_{-0.38}$\\
~~~~$e$\dotfill &Eccentricity \dotfill & $0.298^{+0.022}_{-0.020}$& $0.151^{+0.012}_{-0.011}$\\
~~~~$\tau_{\rm circ}^\pi$\dotfill &Tidal circularization timescale (Gyr)\dotfill & $347^{+100}_{-87}$& $42.1^{+5.1}_{-5.6}$\\
~~~$\omega_*$\dotfill &Argument of Periastron (Degrees)\dotfill & $132.3^{+3.6}_{-3.8}$& $-62.3^{+3.0}_{-2.6}$\\
~~~~$T_{eq}$\dotfill &Equilibrium temperature (K)\dotfill & $1061^{+13}_{-12}$& $1180^{+18}_{-16}$\\
~~~~$K$\dotfill &RV semi-amplitude (m/s)\dotfill & $257.1\pm6.5$& $633.0^{+7.9}_{-8.4}$\\
~~~~$R_P/R_*$\dotfill &Radius of planet in stellar radii \dotfill & $0.0746^{+0.0013}_{-0.0011}$& $0.09097^{+0.00056}_{-0.00050}$\\
~~~~$a/R_*$\dotfill &Semi-major axis in stellar radii \dotfill& $18.56^{+0.57}_{-0.58}$& $12.61^{+0.28}_{-0.32}$\\
~~~~$Depth$\dotfill &Flux decrement at mid transit \dotfill & $0.00557^{+0.00019}_{-0.00017}$& $0.008276^{+0.00010}_{-0.000090}$\\
~~~~$\tau$\dotfill &Ingress/egress transit duration (days)\dotfill & $0.0385^{+0.0045}_{-0.0034}$& $0.01884^{+0.0011}_{-0.00087}$\\
~~~~$T_{14}$\dotfill &Total transit duration (days)\dotfill & $0.1127^{+0.0020}_{-0.0019}$& $0.21459^{+0.0010}_{-0.00090}$\\
~~~~$b$\dotfill &Transit Impact parameter \dotfill & $0.9073^{+0.0066}_{-0.0067}$& $0.230^{+0.088}_{-0.13}$\\
~~~~$T_{S,14}$\dotfill &Total eclipse duration (days)\dotfill & $0.00\pm0.00$& $0.1659^{+0.0045}_{-0.0043}$\\
~~~~$\rho_P$\dotfill &Density (g\ cm$^{-3}$)\dotfill & $3.50^{+0.43}_{-0.41}$& $5.74^{+0.42}_{-0.46}$\\
~~~~$logg_P$\dotfill &Surface gravity \dotfill & $1.16358088\pm0.00000078$& $0.84409860^{+0.00000031}_{-0.00000032}$\\
~~~~$T_S$\dotfill &Time of eclipse (\bjdtdb)\dotfill & $2458366.38\pm0.15$& $2458408.745^{+0.020}_{-0.021}$\\
~~~~$e\cos{\omega_*}$\dotfill & \dotfill & $-0.200\pm0.016$& $0.0700^{+0.0045}_{-0.0047}$\\
~~~~$e\sin{\omega_*}$\dotfill & \dotfill & $0.221^{+0.024}_{-0.022}$& $-0.133\pm0.013$\\
~~~~$d/R_*$\dotfill &Separation at mid transit \dotfill & $13.85^{+0.73}_{-0.75}$& $14.21^{+0.42}_{-0.44}$\\
\hline\\
\end{tabular}
\begin{flushleft}
  \footnotesize{
    \textbf{\textsc{NOTES:}}
$^\dagger$The initial metallicity is the metallicity of the star when it was formed.\\
$^\ddagger$The Equal Evolutionary Point corresponds to static points in a stars evolutionary history when using the MIST isochrones and can be a proxy for age. See \S2 in \citet{Dotter:2016} for a more detailed description of EEP.\\
$^\star$Optimal time of conjunction minimizes the covariance between $T_C$ and Period.\\
$^\star\star\star$The tidal quality factor (Q$_S$) is assumed to be 10$^6$ and is calculated using Equation 2 from \citet{Adams:2006}.\\
In our analysis, we assume the TESS correction for blending should be better than 10\%. Therefore, we adopt a 10\% prior on the blending determined from TICv8 \citep{Stassun:2018_TIC}.
\\
}
 \end{flushleft}
\label{tab:exofast}
\end{table*}

\begin{table*}[!ht]
\centering
\tiny
\caption{Median values and 68\% confidence intervals for the global models}
\begin{tabular}{llcccccccc}
  \hline
  \hline
TOI-558&&\\
\smallskip\\\multicolumn{2}{l}{Wavelength Parameters:}&B&i'&z'&TESS\smallskip\\
~~~~$u_{1}$\dotfill &linear limb-darkening coeff \dotfill &$0.474^{+0.048}_{-0.047}$&$0.206\pm0.048$&$0.169\pm0.046$&$0.225^{+0.028}_{-0.029}$\\
~~~~$u_{2}$\dotfill &quadratic limb-darkening coeff \dotfill &$0.244\pm0.048$&$0.306\pm0.049$&$0.310^{+0.047}_{-0.048}$&$0.318\pm0.028$\\
~~~~$A_D$\dotfill &Dilution from neighboring stars \dotfill &--&--&--&$0.00000\pm0.00032$\\
\smallskip\\\multicolumn{2}{l}{Telescope Parameters:}&PFS\smallskip\\
~~~~$\gamma_{\rm rel}$\dotfill &Relative RV Offset (m/s)\dotfill &$-59.5^{+4.8}_{-4.6}$\\
~~~~$\sigma_J$\dotfill &RV Jitter (m/s)\dotfill &$15.4^{+5.6}_{-3.8}$\\
~~~~$\sigma_J^2$\dotfill &RV Jitter Variance \dotfill &$240^{+200}_{-100}$\\
\smallskip\\\multicolumn{2}{l}{Transit Parameters:}&\TESS\ &\TESS\ &\TESS\ &LCOSAAO (z')&LCOCTIO (B)&LCOCTIO (i')
\\\multicolumn{2}{l}{}& Sector 2 & Sector 3 & Sectors 29+30 & UT 2019-09-28 & UT 2019-10-26 &UT 2020-11-06
\smallskip\\
~~~~$\sigma^{2}$\dotfill &Added Variance \dotfill &$0.000000001^{+0.000000056}_{-0.000000040}$&$0.000000126^{+0.000000091}_{-0.000000064}$&$-0.000000166^{+0.000000099}_{-0.000000095}$&$0.00000032^{+0.00000014}_{-0.00000012}$&$0.00000162^{+0.00000028}_{-0.00000025}$&$0.00000098^{+0.00000023}_{-0.00000020}$\\
~~~~$F_0$\dotfill &Baseline flux \dotfill &$1.000105\pm0.000077$&$0.999951^{+0.000094}_{-0.000095}$&$1.000058^{+0.000046}_{-0.000047}$&$1.000034^{+0.000090}_{-0.000089}$&$1.00031\pm0.00012$&$1.00248\pm0.00014$\\
~~~~$C_{0}$\dotfill &Additive detrending coeff \dotfill &--&--&--&$0.00025^{+0.00025}_{-0.00024}$&$-0.00410\pm0.00027$&$0.00151^{+0.00028}_{-0.00029}$\\
~~~~$C_{1}$\dotfill &Additive detrending coeff \dotfill &--&--&--&$0.00068^{+0.00022}_{-0.00021}$&--&--\\
\hline
TOI-559&&\\
\smallskip\\\multicolumn{2}{l}{Wavelength Parameters:}&R&i'&z'&TESS\smallskip\\
~~~~$u_{1}$\dotfill &linear limb-darkening coeff \dotfill &$0.291\pm0.049$&$0.269\pm0.035$&$0.189\pm0.040$&$0.287^{+0.023}_{-0.024}$\\
~~~~$u_{2}$\dotfill &quadratic limb-darkening coeff \dotfill &$0.273^{+0.048}_{-0.049}$&$0.290^{+0.035}_{-0.034}$&$0.269^{+0.045}_{-0.046}$&$0.284^{+0.028}_{-0.027}$\\
~~~~$A_D$\dotfill &Dilution from neighboring stars \dotfill &--&--&--&$0.00000\pm0.00010$\\
\smallskip\\\multicolumn{2}{l}{Telescope Parameters:}&CHIRON\smallskip\\
~~~~$\gamma_{\rm rel}$\dotfill &Relative RV Offset$^{4}$ (m/s)\dotfill &$-15725.8\pm6.3$\\
~~~~$\sigma_J$\dotfill &RV Jitter (m/s)\dotfill &$13.6^{+7.6}_{-8.2}$\\
~~~~$\sigma_J^2$\dotfill &RV Jitter Variance \dotfill &$180^{+270}_{-150}$\\
\smallskip\\\multicolumn{2}{l}{Transit Parameters:}&Sector 4 01 &Sector 4  O2&Sector 31 \smallskip\\
~~~~$\sigma^{2}$\dotfill &Added Variance \dotfill &$0.000000030^{+0.000000034}_{-0.000000026}$&$0.000000003^{+0.000000018}_{-0.000000015}$&$0.000000006^{+0.000000052}_{-0.000000050}$\\
~~~~$F_0$\dotfill &Baseline flux \dotfill &$1.000008^{+0.000055}_{-0.000054}$&$1.000008\pm0.000035$&$1.000033^{+0.000029}_{-0.000030}$\\
&&PEST UT 2019-09-27 (R)&LCOSSO UT 2019-10-18 (z')&LCOSSO UT 2020-08-20 (i')&LCOSSO UT 2020-08-27 (i')\smallskip\\
~~~~$\sigma^{2}$\dotfill &Added Variance \dotfill&$0.00000727^{+0.00000097}_{-0.00000086}$&$0.00000066^{+0.00000020}_{-0.00000019}$&$0.0000064^{+0.0000013}_{-0.0000011}$&$0.00000069^{+0.00000014}_{-0.00000012}$\\
~~~~$F_0$\dotfill &Baseline flux \dotfill &$1.00291\pm0.00019$&$1.000015\pm0.000095$&$1.00007\pm0.00031$&$0.999998^{+0.000083}_{-0.000082}$\\

~~~~$C_{0}$\dotfill &Additive detrending coeff \dotfill --&$-0.00054^{+0.00029}_{-0.00028}$&--&$-0.00031^{+0.00022}_{-0.00021}$\\
~~~~$C_{1}$\dotfill &Additive detrending coeff \dotfill &--&--&$0.00008^{+0.00058}_{-0.00057}$\\
\hline
\label{tab:exofast_other}
\end{tabular}
\end{table*}

\section{Discussion}
\label{sec:discussion}
Our global model shows that TOI-558 is an F-type star with a mass of $1.349^{+0.064}_{-0.065}$ \msun\ and a radius of $1.496^{+0.042}_{-0.040}$ \rsun. TOI-558 b is a $3.61\pm0.15$ \mj\ planet in a 14.57-day orbit with an eccentricity of $0.298^{+0.022}_{-0.020}$. We characterize TOI-559 as a G dwarf with a stellar mass of $1.026\pm0.057$ \msun and radius $1.233^{+0.028}_{-0.026}$ \rsun; TOI-559 b is $6.01^{+0.24}_{-0.23}$ \mj\, and its orbital period is 6.98 days with an eccentricity of $0.151^{+0.012}_{-0.011}$. Although both planets' masses are likely consistent with core accretion, the mass for TOI-559 b is near the theoretical lower limit for disk fragmentation \citep{Moe:2019}.

We note that we detect a significant long-term RV trend in the multi-year radial velocities of TOI-559. The trend is well fit by a linear velocity variation at a rate of 0.65 m/day. Assuming a circularly bound orbit for the companion, such a trend would correspond to a substellar mass companion with a semi-major axis less than $\sim 8$\,AU, or a stellar massed companion further out. Given the lack of a detected companion in our high spatial resolution observations, stellar companions with separations of $>20$\,AU are unlikely, as they would need to be of significant mass, and therefore luminosity, to induce our observed trend. TOI-559 is worthy of long-term RV monitoring to unveil the nature of its companion.

\begin{figure*}[!ht]
\centering
\includegraphics[width=\linewidth]{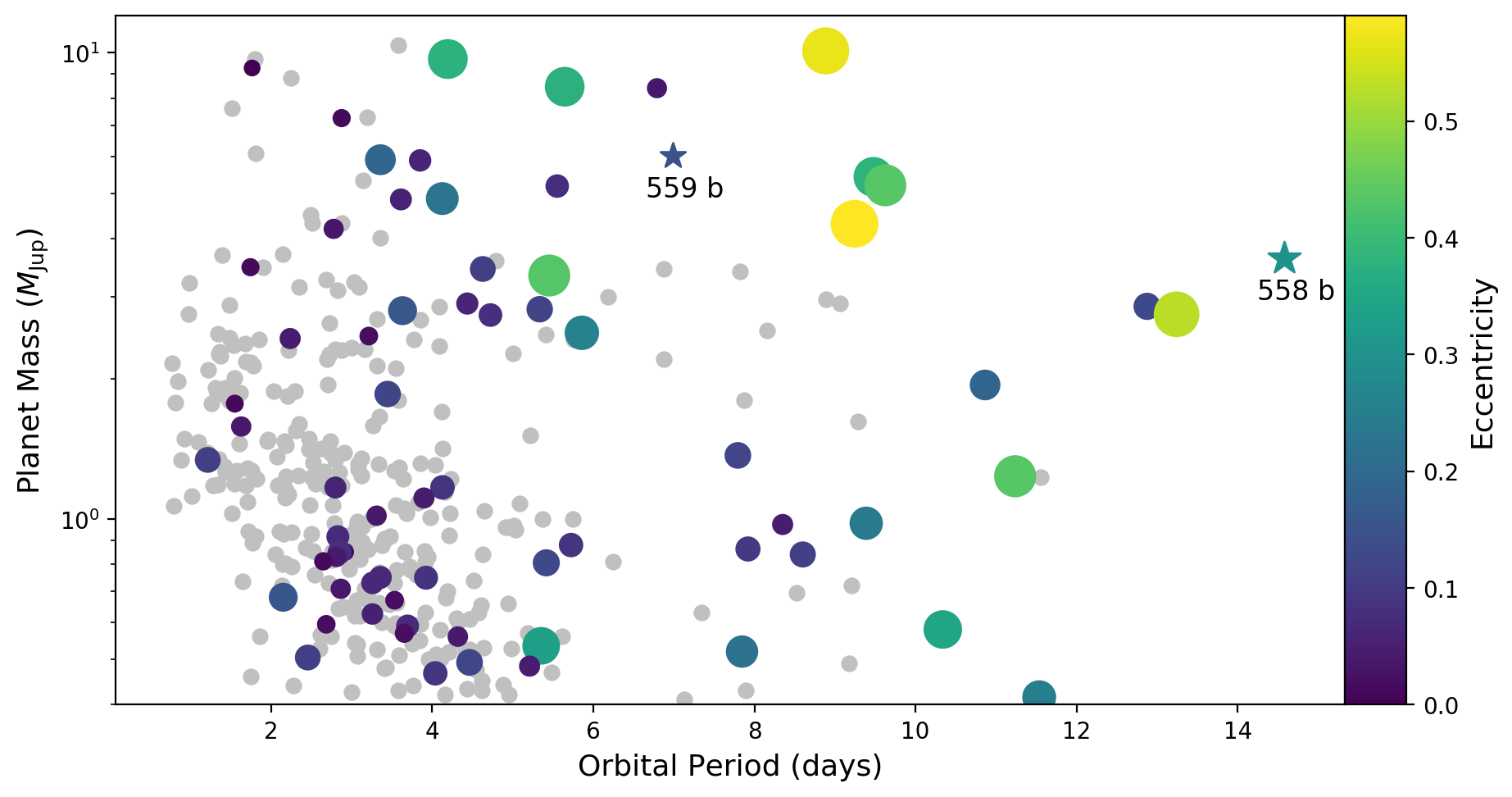}
\caption{The population of transiting giant planets with periods less than 15 days and mass > 0.4 \mj, shown as a function of orbital period versus planet mass, as of UT 1 November 2020. Color and size indicate 1-sigma detection of orbital eccentricity; planets shown in gray do not have significant eccentricity.}
\label{fig:periodmass}
\end{figure*}

\par With high planetary masses and significant orbital eccentricities, TOI-558 b and TOI-559 b occupy a parameter space with few known planets. Only around two dozen previously confirmed transiting giant planets with periods between 5 and 15 days show eccentricity that differs from zero by more than 1 sigma (see Figure \ref{fig:periodmass})\footnote{\label{note1}as of UT 1 November 2020, \url{https://exoplanetarchive.ipac.caltech.edu/}}. Most ground-based surveys have had poor completeness for planets with periods longer than 5 days \citep{Gaudi:2005}, though \tess, which has near-complete sensitivity to hot Jupiters across the main-sequence \citep{Zhou:2019}, will yield many more discoveries in this parameter space. In addition to being particularly massive, TOI-558 b and TOI-559 b have relatively high orbital eccentricities (0.3 and 0.15), indicating that these planets may have migrated to their current orbits through dynamical interactions. Based on the ages of the host stars and our estimates of their respective tidal circularization timescales (see Table \ref{tab:exofast}$^\pi$), we expect that neither of these systems has had sufficient time to circularize. 

\subsection{Period-Mass Distribution}

\begin{figure}[!ht]
\centering
    \includegraphics[width=\linewidth]{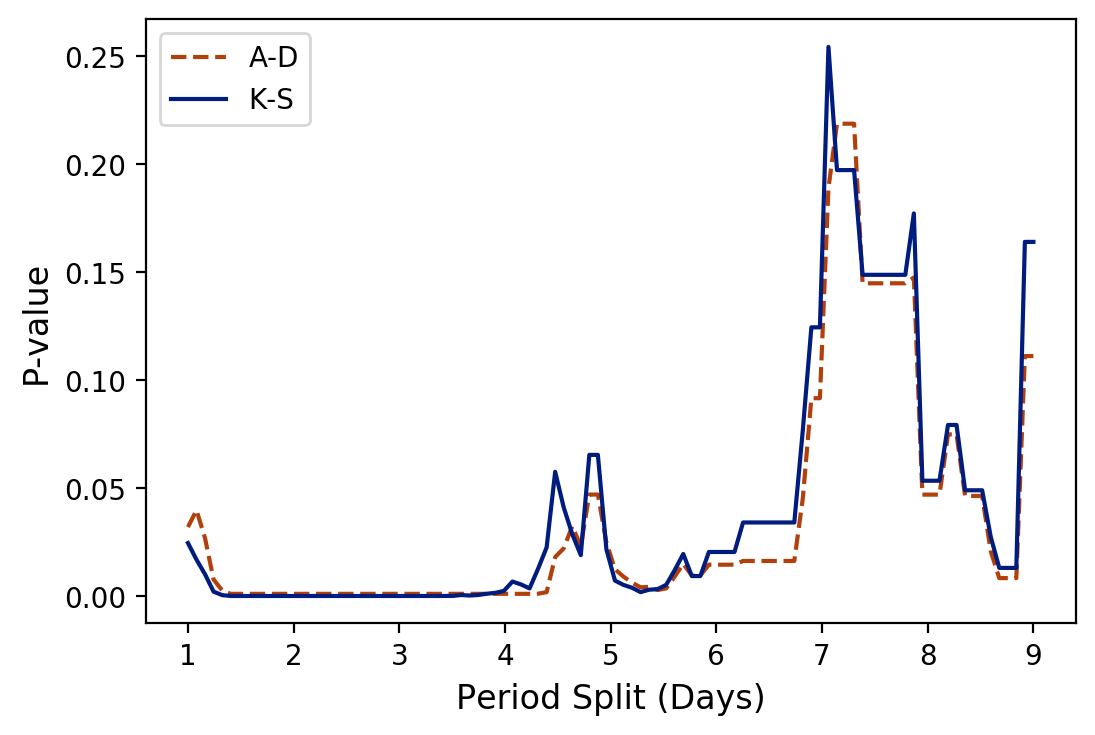}
    \caption{The two-sample Kolmogorov-Smirnov (K-S) and Anderson-Darling (A-D) tests applied to the short-period ($P$<10 days) giant planet ($M_p>0.4 \mj$) population split at orbital periods ranging from 1 to 9 days. The x-axis is the period at which the population is split into two samples, and the y-axis is the resulting p-value. Tidal forces influence the distribution at short periods, possibly shaping the broad minimum between $\sim$ 1.5-4 days, while the minimum at $\sim$ 9 days is likely due to the small sample size at longer periods. The remaining minimum at $\sim$ 5.3 days has no obvious explanation, not showing significant dependence on the lower mass limit chosen for the sample, and could indicate a true break between two distributions.}
    \label{fig:ksrange}
\end{figure}

\begin{figure}[!ht]
\centering
    \includegraphics[width=\linewidth]{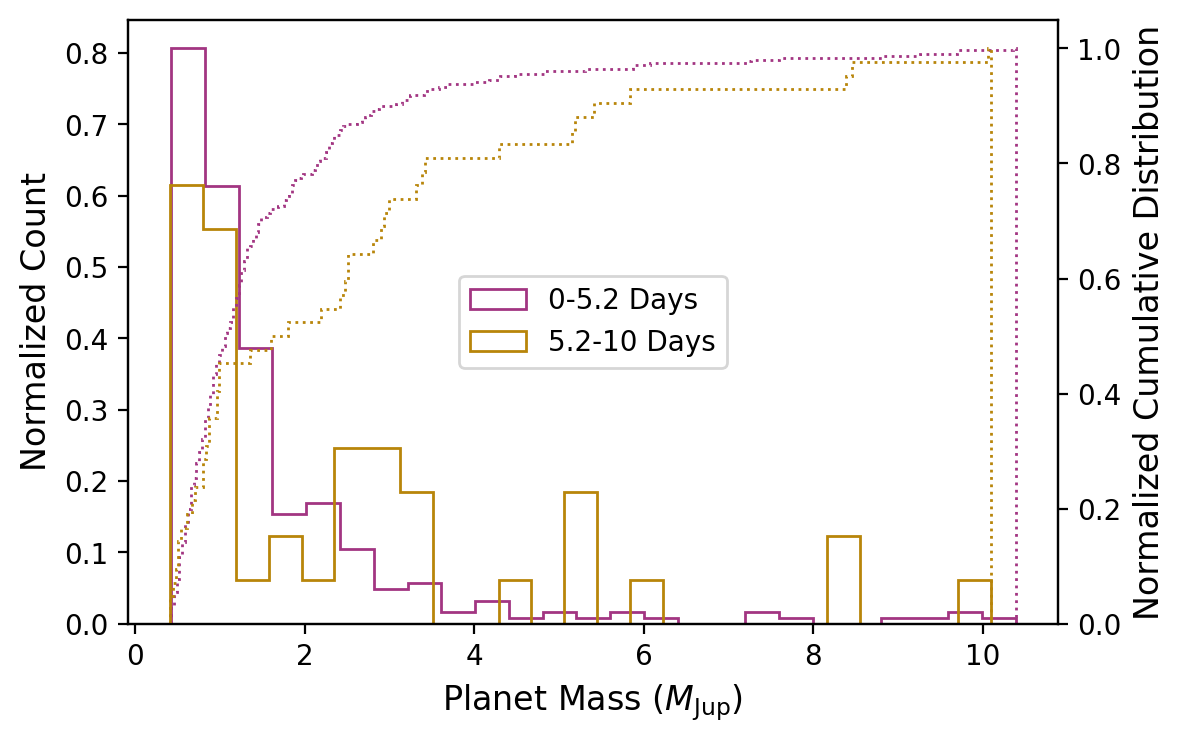}
    \caption{The mass distributions (solid) and cumulative mass distributions (dashed) of all known hot Jupiters with measured masses >0.4 \mj, split at the position of the K-S p-value valley from Figure \ref{fig:ksrange} (P$\approx$ 5.2 days) into two samples. We include 311 planets with periods less than $\sim$ 5.2 days, and 42 with periods between $\sim$ 5.2 and 10 days.}
    \label{fig:distributions}
\end{figure}

{ Like eccentricity, the masses of hot Jupiters may provide clues to their evolutionary processes. For example, if hot Jupiters form in-situ then it is predicted that there may be a a$\sim$M$_P^{-2/7}$ relationship that can be observed in the distribution of planet parameters \citep{Bailey:2018}.} The known population of transiting giant planets with reported masses greater than 0.4 Jupiter masses and orbital periods <15 days is shown in Figure \ref{fig:periodmass} (we exclude planets that do not have reported uncertainties on the mass in the NASA Exoplanet Archive). With \tess\ expected to eventually be magnitude-limited for all transiting hot Jupiters ($P$<10 days), we can test whether possible trends may already exist in the mass distribution of hot Jupiters \citep{Rodriguez:2019b}. To probe this question, we include TOI-558 b and TOI-559 b in a study of the known population of hot Jupiters with periods shorter than 10 days, evaluating the potential existence of multiple populations. We use the \textit{Scipy} implementations of the two-sample Kolmogorov-Smirnov (K-S) test \citep{Massey:1951, Grover:1977} and a two-sample Anderson-Darling (A-D) test \citep{Scholz:1987} to qualitatively identify possible splits in the total population. Across a range of orbital period values, we divide the population into two samples, one with periods shorter than the given value and one with periods longer, and apply the K-S and A-D tests to those two distributions. As shown in Figure \ref{fig:ksrange} \& \ref{fig:distributions}, we find a minimum p-value when the population split occurs between 5 and 5.5 days, at roughly 5.2 days with the K-S test and 5.4 days with the A-D test. In order to limit the influence of detection bias against lower-mass giant planets at longer periods, we include only transiting planets and only those with masses greater than 0.4 \mj\ (with reported mass uncertainties). Given the presence of detection biases at long periods and low masses, it is possible that our sample selection criteria ($M_p > 0.4 \mj$) could affect the result. We therefore rerun the test using 0.3 and 0.5 \mj\ for the minimum mass cutoff for the sample, but we find no qualitative change in the location of the minimum p-value. The presence of this p-value valley may suggest that there are two distributions separated near 5.2 days drawn from distinct parent distributions. The short-period (with 311 planets) and long-period (with 42 planets) samples have mean masses and standard errors of approximately \(1.59 \pm 0.09\) \mj\ and \(2.31 \pm 0.32\) \mj, respectively. The mass distributions and cumulative mass distributions of the two samples are shown in Figure \ref{fig:distributions}.

We caution that the current population of hot giant planets is a heterogeneous sample that comes from a variety of surveys. There are a number of possible biases in the present sample. For example, ground-based surveys have yielded fewer discoveries at longer periods, and there may also be a detection bias against the lowest masses among them. { Physical factors, including the effect of tidal evolution on short-period planets \citep{Jackson:2009}, influence the primordial mass-period distribution. Specifically, these physical factors result in observed features in the population, like the Neptune desert that may extend out to 10 days but is clearly deficient in planets inside of $\sim$2.5 days (See Figure \ref{fig:distributions}, \citealp{Szabo:2011}). } Even for an unbiased sample, it is also possible that an apparent minimum in the p-value like the one we observe could be equally well described by a single, continuous model \citep[e.g.][]{Schlaufman:2015}. Future investigation is warranted, as a more careful characterization of the population may provide constraints on hot Jupiter formation channels. The presence (and characteristics) of two separate hot Jupiter populations---or a single, continuous relationship---in the mass-period plane could be compared to model predictions and simulations of different formation processes and migration mechanisms. There are many confounding variables to consider, such as host star properties, metallicity, system architectures, likely disk conditions, and more, all of which may affect planetary properties and the efficiency of migration mechanisms, and in turn, the expected resulting mass-period distribution. Simply identifying the broad characteristics of the population in mass-period space will require additional discoveries, so a large ensemble--like the complete transiting sample from \tess--will likely be required to draw firm conclusions. TOI-558 b and TOI-559 b represent two examples of planets that can contribute to these types of investigations.

\section{Conclusion}
\label{sec:conclusion}
We present the discovery and detailed characterization of two short-period massive giant planets from the \tess\ FFIs. Globally modeling photometric and spectroscopic observations from \tess\ and ground-based facilities using EXOFASTv2, we confirm TOI-558 b as a $3.62\pm0.15$ \mj\ planet in a $14.574076\pm0.000025$-day orbit around an F-type star, and TOI-559 b to be a $6.01^{+0.24}_{-0.23}$ \mj\ planet in $6.9839115^{+0.0000094}_{-0.0000093}$-day orbit around an early G-dwarf. Additionally, both planets are on eccentric orbits, (e = $0.298^{+0.022}_{-0.020}$ for TOI-558 b and $0.151^{+0.012}_{-0.011}$ for TOI-559 b). The measured eccentricities may be remnant from their evolutionary history since tidal forces at these periods would not have had enough time to circularize their orbits.  A long-term RV trend suggests the presence of an exterior companion to TOI-559, which we do not detect in high-resolution ($\sim 0.1$ arcsecond) images down to limiting contrast of $>5$ magnitudes in the red-optical. Future efforts should continue RV monitoring to constrain the mass and separation of the stellar or substellar companion. The high mass of both planets is also interesting, and we examine the mass distribution of the current known sample of transiting hot and warm Jupiters. While some tentative trends may be present, further work is warranted. Fortunately, \tess\ will provide a near magnitude-complete sample of transiting hot Jupiters \citep{Zhou:2019}, enabling more robust future studies of the population, possibly yielding signatures of migration. Such future work may help illuminate the evolutionary pathways of hot and warm Jupiters, a question that has persisted since the first exoplanet discoveries.

\software{EXOFASTv2 \citep{Eastman:2013, Eastman:2017}, AstroImageJ \citep{Collins:2017}}, Lightkurve \citep{Lightkurve:2018}, Tesscut \citep{Brasseur:2019}, Keplerspline \citep{Vanderburg:2014, Shallue:2018}
\facilities{\tess, FLWO 1.5m (Tillinghast Reflector Echelle Spectrograph), 4.1-m Southern Astrophysical Research (SOAR), LCOGT 0.4m, LCOGT 1.0m, 6.5m Magellan Telescope}

\acknowledgements

This research has made use of SAO/NASA's Astrophysics Data System Bibliographic Services. This research has made use of the SIMBAD database, operated at CDS, Strasbourg, France. This work has made use of data from the European Space Agency (ESA) mission {\it Gaia} (\url{https://www.cosmos.esa.int/gaia}), processed by the {\it Gaia} Data Processing and Analysis Consortium (DPAC, \url{https://www.cosmos.esa.int/web/gaia/dpac/consortium}). Funding for the DPAC has been provided by national institutions, in particular the institutions participating in the {\it Gaia} Multilateral Agreement. This work makes use of observations from the LCOGT network. B.S.G. was supported by a Thomas Jefferson Grant for Space Exploration from the Ohio State State University.

Funding for the \tess\ mission is provided by NASA's Science Mission directorate. We acknowledge the use of public \tess\ Alert data from pipelines at the \tess\ Science Office and at the \tess\ Science Processing Operations Center. This research has made use of the NASA Exoplanet Archive and the Exoplanet Follow-up Observation Program website, which are operated by the California Institute of Technology, under contract with the National Aeronautics and Space Administration under the Exoplanet Exploration Program. This paper includes data collected by the \tess\ mission, which are publicly available from the Mikulski Archive for Space Telescopes (MAST). Resources supporting this work were provided by the NASA High-End Computing (HEC) Program through the NASA Advanced Supercomputing (NAS) Division at Ames Research Center for the production of the SPOC data products. Part of this research was carried out at the Jet Propulsion Laboratory, California Institute of Technology, under a contract with NASA.

This paper includes data gathered with the 6.5 meter Magellan Telescopes located at Las Campanas Observatory, Chile. This paper includes observations obtained under Gemini programs GN-2018B-LP-101 and GN-2020B-LP-105. Some of the observations in the paper made use of the High-Resolution Imaging instrument `Alopeke. `Alopeke was funded by the NASA Exoplanet Exploration Program and built at the NASA Ames Research Center by Steve B. Howell, Nic Scott, Elliott P. Horch, and Emmett Quigley. `Alopeke was mounted on the Gemini North telescope of the international Gemini Observatory, a program of NSF’s OIR Lab, which is managed by the Association of Universities for Research in Astronomy (AURA) under a cooperative agreement with the National Science Foundation. on behalf of the Gemini partnership: the National Science Foundation (United States), National Research Council (Canada), Agencia Nacional de Investigaci\'{o}n y Desarrollo (Chile), Ministerio de Ciencia, Tecnolog\'{i}a e Innovaci\'{o}n (Argentina), Minist\'{e}rio da Ci\^{e}ncia, Tecnologia, Inova\c{c}\~{o}es e Comunica\c{c}\~{o}es (Brazil), and Korea Astronomy and Space Science Institute (Republic of Korea).

\bibliographystyle{style/apj}
\bibliography{refs_joey}

\end{document}